\documentclass[twocolumn,aps,prc,tightenlines,floats,floatfix,nofootinbib]{revtex4}

\def\etal{{\it et.\ al.\/},$\;$}

\def\bea{\begin{eqnarray}}
\def\eea{\end{eqnarray}}
\def\etal{{\it et al.\/}}

\def\etal{{\it et al.\/}}
\def\sfrac#1#2{{\textstyle \frac{#1}{#2}}}

\newcommand{\bra}[1]{\langle #1|}
\newcommand{\ket}[1]{|#1\rangle}
\newcommand{\braket}[2]{\langle #1|#2\rangle}

\def\be{\begin{equation}}
\def\ee{\end{equation}}
\def\ba{\begin{eqnarray}}
\def\ea{\end{eqnarray}}

\usepackage{graphics}
\usepackage{graphicx}
\usepackage{epsf}
\usepackage{amsmath}
\usepackage{amssymb}

\begin{document}

\phantom{0}
\vspace{-0.2in}
\hspace{5.5in}
\parbox{1.5in}{ } 

\vspace{-1in}

\title
{\bf A covariant model for the 
$\gamma N \to N(1535)$ transition at high momentum transfer}

\author{G.~Ramalho$^1$ and M.~T.~Pe\~na$^{1,2}$ \vspace{-0.15in}}

\affiliation{
$^1$CFTP, Instituto Superior T\'ecnico,
Av.\ Rovisco Pais, 1049-001 Lisboa, Portugal \vspace{-0.15in}}

\affiliation{
$^2$Dep.~F{\'\i}sica, Instituto Superior T\'ecnico,
 Av.\ Rovisco Pais, 1049-001 Lisboa, Portugal}

\vspace{0.2in}
\date{\today}

\phantom{0}

\begin{abstract}
A relativistic constituent quark model is applied to 
the $\gamma N \to N(1535)$ transition.
The $N(1535)$
wave function is determined by extending the 
covariant spectator quark model,  previously developed for the nucleon, to the 
$S_{11}$ resonance.
The model allows us  to calculate the 
valence quark contributions to the 
 $\gamma N \to N(1535)$ transition form factors.
Because of the nucleon and  $N(1535)$ structure 
the model is valid only for $Q^2> 2.3$ GeV$^2$.
The results are compared with the 
experimental data for the electromagnetic form factors 
$F_1^\ast$ and $F_2^\ast$ and 
the helicity amplitudes $A_{1/2}$ and 
$S_{1/2}$,  at high $Q^2$.
\end{abstract}

\vspace*{0.9in}  
\maketitle

\section{Introduction}

The quark and gluon substructure of the hadrons is 
ruled by quantum chromodynamics (QCD), and  it
is reflected in the baryon sector 
by a set of bumps in the cross sections of different probing processes, taken
as functions of the center of mass energy $W$.
These bumps are identified as baryon resonances 
characterized by
spin, isospin, orbital angular momentum,
radial excitation and parity quantum numbers. 
The lowest energy bump, the $\Delta(1232)$ baryon, 
is clearly isolated from 
the background as a state of spin and isospin 3/2 
and positive parity. 
Heavier resonances are not so clearly isolated 
from the background. This happens
in the so called second resonance region, where the 
$P_{11}(1440)$, $D_{13}(1520)$ and
$S_{11}(1535)$ resonances show up.
Although in quark models these resonances 
can be described as three-quark systems 
confined by a potential 
like the harmonic oscillator potential
\cite{Giannini91,Glozman96,Capstick00,Isgur77,Isgur79}, 
some properties, like their decay width, can be better 
understood within a  dynamical meson-baryon 
coupled-channel reaction model.
Also, in constituent quark models the baryon spectrum is difficult to
interpret since the 
negative parity partner of the nucleon, 
the $S_{11}$ state ($J^P=\frac{1}{2}^-$)  
is  lighter 
than the first radial excitation of 
the nucleon  ($J^P=\frac{1}{2}^+$), the Roper 
(or $P_{11}$ state) \cite{Capstick00,Mathur05}.   
It was only recently that lattice QCD simulations 
with very small pion masses \cite{Mathur05},
reconstructed the natural order 
of the baryon spectrum 
(where the $S_{11}$ state is heavier than the $P_{11}$ state),
suggesting a fundamental role 
of the quark-antiquark polarization,
or meson cloud dressing, 
in the baryon systems, as a correction 
to the valence quark effects. 

In this work we will use the notation $N(1535)$ 
to represent the $S_{11}(1535)$ nucleon excitation ($N$), and
we will focus on  the electromagnetic structure 
of this resonance, 
in particular on the calculation of the $\gamma N \to N(1535)$ transition 
form factors, within a covariant constituent quark model.
Precise data for the  $\gamma N \to N(1535)$ 
amplitudes is available at present
\cite{Aznauryan09,MAID,Dalton09,Brasse84,Armstrong99,Thompson01,Denizli07,Burkert04}. 
Besides being one of the lightest nucleon 
resonances, the  $N(1535)$ baryon is particularly interesting 
for several reasons: it is very well isolated in the spin 1/2 and 
negative parity configuration; 
it decays strongly to the $\eta N$ channel 
(with a branching ratio $\approx $ 50\%), 
allowing a very precise determination 
of the electromagnetic structure,  and
providing therefore an extra challenge for theoretical models.
Also, because of the strong coupling 
with the $\pi N$ channel  
(with a branching ratio $\approx$ 50\%), 
the 
$N(1535)$ is crucial  for the analysis 
of meson photoproduction from the nucleon 
\cite{Burkert04}.
Another interesting aspect of the $N(1535)$
is its vicinity to another $S_{11}$ 
resonance with higher mass, the $S_{11}(1650)$ also called as $N(1650)$.
The two resonances differ in their decay modes, 
and the differences in their structure is yet to be explored.

Several formalisms have been used 
to describe the  
$N(1535)$ system. They are
based either on quark models 
or on effective meson-baryon interaction models. In the first case,
there are non-relativistic 
constituent quark models \cite{Isgur77,Isgur79,Warns90,Close90,Li90,Capstick95,Aiello98,Aznauryan08,Zhao02,Golli11}, relativistic 
quark models \cite{Konen90,Capstick95,Pace,He06},
quark models with explicit quark-antiquark contributions \cite{An08}
and QCD sum rules \cite{Braun09}.
Alternatively, in the second case, the $N(1535)$
is interpreted as a molecular-type state
dynamically generated by the meson-nucleon interaction 
\cite{Kaiser95,Li96,Nieves01,Inoue01,Chiang03,Hyodo08,Jido08,Doring09b,Oset10,Bruns10}
with a particular dominance of the $K \Sigma$ 
quasi-bound state \cite{Kaiser95,Li96,Chiang03}.
A particular class of effective 
meson-baryon interaction models are 
the dynamical coupled-channel reaction models 
\cite{MAID,Chen03,Doring09a,Julich,Matsuyama07,JDiaz09,Golli11},
where the baryon bare core is parametrized phenomenologically
and the meson dressing is included non perturbatively.
Thus $N(1535)$ does not only provide a crucial test for the 
methods just mentioned, but it is also a crucial resonant structure 
for the analysis of nucleon excitation reactions 
\cite{Burkert04,MAID,Arndt08,Aznauryan09,Bonn,Doring06,Penner02,CMB,KSU,Doring09a,Mokeev09}.

Within a constituent quark model picture, 
the nucleon excitation  $N(1535)$ can be represented 
as a mixture of two different configurations.
Since the $S_{11}$  excitation has total angular momentum $J=1/2$
and orbital angular $L=1$ ($P$ state excitation), 
its core spin may be either  $S=1/2$ or $S=3/2$.
Then, in the usual spectroscopic notation 
\cite{Capstick00,Chiang03}, the  $S_{11}$ channel of the nucleon excitation
is a mixture of the $\ket{N\, ^2P_{1/2}}$  and $\ket{N\, ^4P_{1/2}}$ 
states, which have spin 1/2 and 3/2 respectively.
This mixture of the two core spin  components 
is defined by a mixing angle $\theta_S$ determined by a
color hyperfine interaction 
between the quarks, which may have distinct origins: 
one-gluon-exchange \cite{Capstick00,Isgur77,Isgur79}, 
one-pion-exchange \cite{Chiang03} or
Goldstone-boson-exchange \cite{He03}.
In the classical Isgur-Karl model it turns out that
the spin core spin 1/2 component dominates 
in the $N(1535)$, with a mixing angle given by $\cos \theta_S= 0.85$
\cite{Isgur77,Isgur79}.

In our work, we apply the covariant spectator quark model, which is 
based on  the covariant spectator theory
\cite{Gross}, to the $N(1535)$ system. 
The model describes the nucleon 
\cite{Nucleon,FixedAxis,Lattice,Octet}, the Roper 
\cite{Roper,ExclusiveR}, the $\Delta (1232)$ and the
$\Delta(1600)$ \cite{ExclusiveR,NDelta,NDeltaD,Lattice,LatticeD,DeltaFF0,DeltaDFF,Delta1600} experimental form factors,
as well as the lattice QCD simulations  
for the nucleon, the $\gamma N \to \Delta$ transition,
and the baryon decuplet \cite{Lattice,LatticeD,Omega,GE2Omega}.  
In our framework the baryons are represented 
as a quark-diquark system. 
The quark couples 
to the electromagnetic field  by means of  
a constituent quark current which is parametrized by  
vector meson dominance, 
and the diquark is a spectator 
during 
the electromagnetic interaction, and therefore is taken on-mass-shell
\cite{Nucleon,NDeltaD,Omega,DeltaDFF}.
The model is phenomenological 
since it does not derive 
the structure of the baryon from a dynamical 
wave function equation. Instead,
the baryon systems are described effectively
in terms of their intrinsic properties 
(spin, flavor, angular orbital momentum and parity) --- which dictate the
form of their wave function --- and the experimental value of their
mass $M_B$.
As in the previous applications of the model, in
particular to the $\Delta$ and the Roper resonances, we are focused 
on the role of the valence quarks for the electromagnetic transition.
Because of this and also as a consequence 
of the kinematics 
(the difference of mass between the $N(1535)$
and the nucleon is 0.60 GeV)
our model can only be applied
to the high $Q^2$ region. As we will show, the domain 
of validity of our calculations can even be established
 more precisely and quantitatively,
as the region $Q^2> 2.3$ GeV$^2$.
In this region the meson cloud effects are 
expected to be small and  valence quark degrees to
dominate.
We use two additional assumptions:
i) the $N(1535)$ is represented exclusively by the spin 1/2 core 
[no mixture with the $N(1650)$ excitation]
ii) the diquark is pointlike.
With these assumptions, and taking 
the momentum distribution of the diquark 
the same as for the nucleon,
we relate the  nucleon and the $N(1535)$ wave functions.
These assumptions allow us to reduce 
the number of degrees of freedom to a minimum, 
since no additional parameters to the ones taken 
for the nucleon case are needed 
to describe the spin 3/2 core  
contributions, or the diquark internal structure. 
Our results are then true predictions,
with no new adjustable parameters.
All parameters were fixed in the previous 
applications by the quark current and nucleon 
wave function, represented as S-wave system.
Both assumptions can be tested 
in the future, once the structure of the nucleon 
is extended to  the inclusion of P- and 
D-states, which demand in turn a spin 3/2 core and/or diquark with internal 
P-state structure \cite{InProgress}.

This work will be organized as follows:
In Sec.~\ref{secSQM} we introduce 
the wave functions of the nucleon 
and the $N(1535)$ (details in Appendix \ref{apWF}).
In Sec.~\ref{secCurrent} we derive the 
transition current for the
 $\gamma N \to N(1535)$  transition 
(with details presented in Appendix \ref{apCurrent}).
Explicit formulae for the form factors 
and helicity amplitudes come 
in Sec.~\ref{secFF}.
In Sec.~\ref{secWF} we parametrize 
the momentum dependence of the wave functions.
The results and discussion are
presented in Sec.~\ref{secResults}
and the conclusions in Sec.~\ref{secConclusions}.

\section{Spectator quark model}
\label{secSQM}
When the momentum transfer  exceeds the mass of the constituent quarks
the electromagnetic excitation requires necessarily a relativistic treatment. This is one of the reasons for us to
use the framework provided by the covariant spectator quark model
for baryons \cite{Nucleon}. In this formalism the baryons 
are phenomenologically 
described as constituent quark systems, and the
covariant wave function has a form compatible with 
their symmetry properties 
(flavor, spin, orbital angular momentum and parity) 
and a totally anti-symmetric color wave function
\cite{Nucleon,NDeltaD,Roper,Omega}.

\subsection{Nucleon wave function}

For the nucleon  the S-state 
approximation was made and the spin, flavor and spatial wave function
and is represented
by \cite{Nucleon}
\be
\Psi_N(P,k)=
\frac{1}{\sqrt{2}} \left[
\phi_I^0 u(P) - \phi_I^1 
\left(\varepsilon_P^\ast\right)_\alpha U^\alpha(P)
\right] \psi_N(P,k),
\label{eqPsiN}
\ee 
where the nucleon  and diquark four momenta are $P$ and $k$ 
respectively,   $u$ is a Dirac spinor, 
$\varepsilon_P$ the diquark polarization vector 
in the fixed-axis representation \cite{FixedAxis} and 
\be
U^\alpha(P)= \frac{1}{\sqrt{3}} 
\gamma_5 \left(\gamma^\alpha -\frac{P^\alpha}{M} \right)u(P),
\label{eqUN}
\ee
the spin 1/2 vector spin state 
[direct product of states 1 (diquark) 
and 1/2 (quark) for a total spin state of 1/2].
$M$ is the nucleon mass.
The wave function (\ref{eqPsiN}) is written in terms 
of the states corresponding to a diquark 
composed by the quark pair (12) and the quark 3.
The isospin functions $\phi_I^{0,1}$ depend on
the isospin projection $\pm 1/2$ and are shown in Table 
\ref{tabIsospin}.   
Note that the spin-0  (isospin-0)  and the spin-1 (isospin-1) states
are respectively anti-symmetric and symmetric 
in the exchange of quarks 1 and 2.

\begin{table}[t]
\begin{center}
\begin{tabular}{lcc}
\hline
\hline
 & $\phi_I^0$  & $\phi_I^1$\\
\hline
$p$ & $\qquad \frac{1}{\sqrt{2}}(ud -du)u \qquad $ & 
$ \frac{1}{\sqrt{6}}\left[(ud +du)u -2 uud \right] $ \\
$n$ & $\qquad \frac{1}{\sqrt{2}}(ud -du)d \qquad$ & 
$\frac{1}{\sqrt{6}}\left[2 dd u- (ud +du)d  \right]$ \\
\hline
\hline
\end{tabular}
\end{center}
\caption{Isospin states for the nucleon and $S_{11}$ systems.}
\label{tabIsospin}
\end{table}

\subsection{$N(1535)$ wave function}

To write down the $N(1535)$ wave function we applied the  
$SU(3) \otimes O(3)$ constituent quark model 
representation, where the $N(1535)$ state is a member 
of the $[70,1^-]$ supermultiplet 
(dimension 70, with $L^P=1^-$), and part
of the $^28$ subset (octet with $2S+1=2$)
\cite{Giannini91,Capstick00,Isgur77,Isgur79,Warns90,Close90,Li90}. 
We have also  followed very closely the notation established in 
Refs.~\cite{Capstick00,Burkert04,Hey74,Cottingham78}. 
We use $M_S$ to label the $N(1535)$ mass.

The $N(1535)$ is defined as the excitation of the nucleon 
to the state $I(J^P)=\sfrac{1}{2}\left(\sfrac{1}{2}\right)^-$.
This state has the same flavor content and the same spin  (1/2) 
of the nucleon,  but has negative parity. 
The negative parity
defines a spatial symmetry implied by the excitation of internal  
relative angular momentum
$L=1$, and requires the presence of P waves at
least in one quark pair. 
Consequently,
the spin structure also changes relatively to the one 
of the nucleon, in order to accommodate
a total symmetric form for the 
flavor-spin-momentum space wave function.

To represent the wave function in a basis of momentum states, one 
decomposes, as usual, the system 
into a pair of quarks [or diquark labeled (12)],  and a spectator quark 
[labeled quark (3)], 
and one defines the momentum variables corresponding to
those diquark and spectator quark sub-systems 
(the  so-called Jacobi momenta). 
If the individual quark momenta are $k_i$ ($i=1,2,3$), 
the Jacobi momenta are $k_\rho= \frac{1}{\sqrt{2}}(k_1-k_2)$, 
the relative momentum of the quarks 
in diquark (12),
and $k_\lambda=\frac{1}{\sqrt{6}}(k_1+k_2-2k_3)$,
the diquark center of mass momentum
with with respect to quark (3).
The center of mass momentum is $P=k_1+ k_2+k_3$.
The momentum states that define our basis to represent the 
wave function are the eigenvectors of the Jacobi momenta 
$k_\lambda$ and $k_\rho$.
They are called $\lambda$-type and  $\rho$-type states, 
with mixed symmetry\footnote{The Jacobi momentum
$k_\rho$  is anti-symmetric 
for the exchange of quarks 1 and 2,
while the Jacobi momentum $k_\lambda$ is symmetric for the same exchange. 
The Jacobi momenta $k_\rho$  and $k_\lambda$  eigenvector 
basis states are therefore anti-symmetric and symmetric, 
respectively, under that exchange.
For another  particle exchange  $i \rightarrow j$, with $(ij) \ne (12)$, 
those states are, however, states of mixed symmetry.}.
Following the traditional notation 
(see  e.g.~Ref.~\cite{Capstick95,Aznauryan08,Chen03}), 
the labels $\rho$ and $\lambda$ are
used more generally, i.e., 
for combinations and angular momentum projections of momentum states,
and also for spin and isospin states, that are, respectively, 
anti-symmetric and symmetric 
under the exchange of quarks (12). 

The starting point for the construction of the 
flavor-spin-momentum-space wave function is 
to impose that it is symmetric under the exchange of any
pair (the color part, which is omitted, makes it anti-symmetric 
at the end, as required).
The second step is to write the non relativistic limit 
of the wave function 
in terms of $\lambda$-type and $\rho$-type mixed-symmetric states, 
labeled $X_{\rho}$  and $X_{\lambda}$, that couple orbital states $L=1$ 
(in principle in both $k_\rho$ and $k_{\lambda}$ Jacobi momenta) 
with total three-quark spin $S=1/2$ states, 
and to multiply them with the adequate 
flavor states that make the function symmetric.
Next, we assume a pointlike diquark. 
In this approximation, effectively, one has $k_\rho \equiv 0$.
With this suppression of the diquark internal P states, 
the orbital wave function is reduced  to P-states 
in  the momentum $k_\lambda$ of the quark-diquark motion only.
Additionally, the non relativistic wave function 
is calculated in the 3 body center of mass frame, 
where ${\bf k_1}+ {\bf k_2}+{\bf k_3}={\bf 0}$,
and the diquark three momentum becomes 
${\bf k}= {\bf k_1} + {\bf k_2}=- {\bf k_3}$.
Then, the spin-orbital part of the non-relativistic 
wave function is, in our approximation,
written as a function of ${\bf k}_\lambda= \sqrt{\frac{3}{2}} {\bf k}$ only.

Finally, one makes the relativistic generalization of 
the coupled spin-orbital states $X_\rho$ and $X_\lambda$. 
The  corresponding relativistic states, labeled respectively 
$\Phi_\rho$ and $\Phi_\lambda$,  include a $\gamma_5$ matrix,  
exhibiting the negative parity of the state explicitly.
All the details concerning the full non-relativistic  wave 
function in the pointlike diquark limit, 
and its relativistic generalization,
are presented in Appendix \ref{apWF}.
To conclude this section, we write in the pointlike diquark 
approximation, the final expression for the 
covariant  structure of the spin-flavor-orbital wave function 
of the $N(1535)$. It
depends on the baryon four-momentum $P$  and on the diquark 
four momentum $k$, and  is given by
\be
\Psi_{S11}(P,k)
=\frac{1}{\sqrt{2}}
\left[ 
\phi_I^0 \Phi_\rho  -
\phi_I^1 \Phi_\lambda
\right] \psi_{S11}(P,k),
\label{eqPsiS11}
\ee
where 
$\phi_I^0$ and $\phi_I^1$ are the flavor states,
and 
\ba
& &
\hspace{-0.5cm}
\Phi_\rho(\pm)=
-\gamma_5
N \left[
(\varepsilon_0 \cdot \tilde k) u_S(\pm) -
\sqrt{2} (\varepsilon_\pm \cdot \tilde k) u_S(\mp) 
\right]  \nonumber \\
& &
\hspace{-0.5cm}
\Phi_\lambda(\pm)=
+
\gamma_5 N \left[
(\varepsilon_0 \cdot \tilde k) \varepsilon_\alpha^\ast 
U_S^\alpha (\pm) -
\sqrt{2} (\varepsilon_\pm \cdot \tilde k)  
 \varepsilon_\alpha^\ast 
U_S^\alpha (\mp)
\right]. \nonumber \\
& & 
\ea
In the last equations $\tilde k= k - \sfrac{P \cdot k}{M_S^2}P$
and $N=1/\sqrt{- \tilde k^2}$.
The four momentum $\tilde k$ can 
be interpreted as the diquark three momentum 
in the $N(1535)$ rest frame
[where $\tilde k=(0,{\bf k})$ 
and $\tilde k^2= -{\bf k}^2$].
The spinors $u_S$ and $U_S^\alpha$ 
have the same meaning as $u$ and $U^\alpha$, 
defined for the nucleon before \cite{Nucleon,NDelta,NDeltaD},
as in Eq.~(\ref{eqUN}),    
but are here associated with the $N(1535)$ baryon.

The scalar wave function $ \psi_{S11}(P,k)$ 
will be discussed later (see Sec.~\ref{secWF}).
Here it suffices to say that this function carries all the information 
on the momentum distribution of the quark-diquark relative motion, it is purely 
phenomenological and normalized to one.

We make two more notes about Eq.~(\ref{eqPsiS11}):
The wave function in our model does not contain the
contribution of three-quark states with total spin $S=3/2$,
included in other works \cite{Isgur77,Capstick95,Chiang03}.
Additionally, the minus sign  for the $\lambda$-type spin-orbital in the
wave function is needed to ensure orthogonality
between the $N(1535)$ and the nucleon
wave functions in the non relativistic limit \cite{Chiang03}.

\section{Transition current}
\label{secCurrent}

We can write the transition current in
relativistic impulse approximation 
\cite{Nucleon,Omega} as
\be
J^\mu= 3
\sum_{\Lambda} \int_k \overline \Psi_{S11}(P_+,k) 
j_I^\mu
\Psi_N(P_-,k),
\label{eqJI1}
\ee
where $\Lambda=\left\{s,\lambda_D\right\}$ 
(scalar diquark $s$ and vector diquark polarization $\lambda_D=0,\pm 1$) 
and $\int_k \equiv \int \frac{ d^3 k}{(2\pi)^22E_D}$ is
the covariant integration element 
in the diquark on-mass-shell momentum $k$ 
(mass $m_D$ and energy $E_D$).
The factor 3 accounts for the contributions 
of all possible diquark pairs, since, due 
to the symmetry of the wave function, pairs (13) and (23) 
give the same contribution as pair (12).
[The magnitude of the electron charge $e$ 
was not included in the current for simplicity].
In the previous equation, $j_I^\mu$ is
the quark current
\be
j_I^\mu = j_1 \left(
\gamma^\mu -\frac{\not \! q q^\mu}{q^2}\right)+
j_2 \frac{i \sigma^{\mu \nu} q_\nu}{2M}. 
\ee

To obtain the $\gamma N \to N(1535)$ transition current
we take the wave functions 
(\ref{eqPsiN}) and (\ref {eqPsiS11}).
To work the spin algebra one uses 
$j_i \to   \left(\phi_I^0 \right)^\dagger\! j_i \phi_I^0$
and 
$j_{(i+2)} = \left(\phi_I^1 \right)^\dagger \!j_i \phi_I^1$,
obtaining
\ba
j_i &= & \sfrac{1}{6} f_{i+} + \sfrac{1}{2} f_{1-} \tau_3 \\
j_{(i+2)} &= & \sfrac{1}{6}f_{i+} - \sfrac{1}{6}f_{i-} \tau_3.
\ea
The coefficients $j_{1,2}$ and $j_{3,4}$ follow the definitions in Ref.~\cite{Nucleon}.
Note that the result is a sum over the flavor of 
the anti-symmetric ($j_1$  and $j_2$) 
and symmetric components ($j_3$ and $j_4$) 
as done in Refs.~\cite{Octet,Omega} for the SU(3) case.
For convenience one introduces also the notation
\be
\hat \gamma^\mu= \gamma^\mu - \frac{\not\! q q^\mu}{q^2}.
\ee

Using the definitions above
one can write 
\ba
\sum_{\Lambda}{\overline \Psi}_{S11} j_I^\mu \Psi_N &=&
\frac{\cal A}{2}
\left\{
j_1 \overline \Phi_\rho \hat \gamma^\mu \phi_S^0+  
j_2 \overline \Phi_\rho \frac{i \sigma^{\mu \nu} q_\nu}{2M} \phi_S^0
\right\} \nonumber \\
& -& 
\frac{\cal A}{2}
\left\{
j_3 \overline \Phi_\lambda \hat \gamma^\mu \phi_S^1+  
j_4 \overline \Phi_\lambda \frac{i \sigma^{\mu \nu} q_\nu}{2M} \phi_S^1  
\right\}, \nonumber \\
& &
\label{eqSum}
\ea
where ${\cal A}=\psi_{S11} \psi_N$.
For the vector diquark contributions (terms in $\phi_S^1$) 
the sum in the diquark polarization $\lambda_D$ is implicit.
The isovector components include a sum in the 
diquark polarizations $\lambda_D$ vectors associated with the $N(1535)$,
$\varepsilon_{P_+}^\alpha(\lambda_D)$,  and the nucleon, 
$\varepsilon_{P_-}^{\beta\, \ast}(\lambda_D)$.
Those polarization vectors are functions
of the $N(1535)$ mass ($M_S$) and the nucleon ($M$) mass, respectively 
(see details in Ref.~\cite{FixedAxis} where this basis of states is explained and built).
By adding  the diquark polarizations, one has  
\cite{FixedAxis,NDelta}
\ba
\Delta^{\alpha \beta} & \equiv &\sum_{\lambda_D}  
\varepsilon_{P_+}^\alpha (\lambda_D) 
\varepsilon_{P_-}^{\beta\, \ast} (\lambda_D)  \nonumber \\
&=& -\left(g^{\alpha \beta} - \frac{P_-^\alpha P_-^\beta}{P_+ \cdot P_-}
\right) \nonumber \\
& &- 
a \left( P_-- \frac{P_+ \cdot P_-}{M_S^2} P_+\right)^\alpha 
\left( P_+- \frac{P_+ \cdot P_-}{M^2} P_-\right)^\beta, \nonumber \\
& &
\label{eqDelta}
\ea
where
\be
a= \frac{M_S M}{P_+ \cdot P_-(M_S M +P_+ \cdot P_-)}. 
\ee

The decomposition (\ref{eqSum}) reduces the 
determination of the current (\ref{eqJI1}) to 
the calculation of a few current elements.
The details are presented in Appendix \ref{apCurrent}.
The final result is
\ba
J^\mu &=& \frac{1}{2}(3j_1+ j_3) 
{\cal I}_0 \bar u_S \hat \gamma^\mu  \gamma_5 u \nonumber \\
&-& \frac{1}{2}(3j_2- j_4) 
{\cal I}_0 \bar u_S 
\frac{i \sigma^{\mu \nu} q_\nu}{2M} 
  \gamma_5 u, 
\label{eqJsp}
\ea
where
\be
{\cal I}_0 (Q^2)=
\int_k N (\varepsilon_0 \cdot \tilde k) \psi_{S11} \psi_N.
\label{eqInt0}
\ee
The integral ${\cal I}_0$ is covariant
and includes the dependence of the 
form factors on the  initial and final 
state scalar wave functions.
We call ${\cal I}_0$ the overlap integral.

\section{Form factors and helicity amplitudes}
\label{secFF}

The transition current can be written 
(suppressing the charge factor $e$) as
\cite{Aznauryan08,Braun09}:
\be
J^\mu=
\bar u_S \left[
\left(\gamma^\mu -\frac{\not \! q q^\mu}{q^2}\right) F_1^\ast
+ \frac{i \sigma^{\mu \nu} q_\nu}{M_S+ M} F_2^\ast
\right]\gamma_5 u,
\label{eqJS}
\ee
where $F_i^\ast$ defines the transition form factors.
One should note that there are alternative but equivalent conventions 
for the two form factors \cite{Aznauryan09,Pace,Braun09}.

From the Eqs.~(\ref{eqJsp}) and (\ref{eqJS}), we conclude that
\ba
& &
F_1^\ast(Q^2)= \frac{1}{2}(3j_1+ j_3)  {\cal I}_0 
\label{eqF1}\\
& &
F_2^\ast(Q^2)= -\frac{1}{2}(3j_2- j_4) \frac{M_S+M}{2M} {\cal I}_0 
\label{eqF2}
\ea
 
The experimental data is usually presented 
in terms of the helicity amplitudes in the 
final state (excited resonance) rest frame.
The helicity amplitudes are 
defined from the projection of the current 
on the photon polarization states,
$\epsilon_\lambda^\mu$ and  nucleon and resonance 
spin projections (in the resonance frame).
For a resonance $N^\ast$ with spin 1/2, there are 
two independent amplitudes:
\ba
& &
A_{1/2}(Q^2)= {\cal K} 
\bra{N^\ast,+\sfrac{1}{2}} \varepsilon_+ \cdot J
\ket{N, -\sfrac{1}{2}},
\label{eqA12}\\
& &
S_{1/2}(Q^2)= {\cal K} 
\bra{N^\ast,+\sfrac{1}{2}} \varepsilon_0 \cdot J
\ket{N, + \sfrac{1}{2}} \frac{|{\bf q}|}{Q}.
\label{eqS12}
\ea
Considering $N^\ast=N(1535)$, the multiplicative constant is
\be
{\cal K}= \sqrt{\frac{2\pi \alpha}{K}},
\ee
with $e=\sqrt{4\pi \alpha}$ is the magnitude of the
electron charge with $\alpha \simeq 1/137$,
and 
$K=\frac{M_S^2-M^2}{2M_S}$. 
The variable $|{\bf q}|$ is
the photon three momentum in the excitation, 
in the $N(1535)$ rest frame, 
\be
|{\bf q}|= \frac{\sqrt{Q_+^2Q_-^2}}{2M_S},
\label{eqq2}
\ee
where $Q_\pm^2= (M_S \pm M)^2 + Q^2$, with $Q^2=-q^2$.

The helicity amplitudes can be represented in terms 
of the form factors \cite{Aznauryan08}:
\ba
\hspace{-1cm}
& &
A_{1/2}= 
-2b  
\left[F_1^\ast + \frac{M_S-M}{M_S+M} F_2^\ast
\right]
\label{eqA12X}
\\
\hspace{-1cm}
& &
S_{1/2}= 
\sqrt{2}b(M_S+M) \frac{|{\bf q}|}{Q^2}  
\left[\frac{M_S-M}{M_S+M} F_1^\ast - \tau F_2^\ast
\right],
\label{eqS12X}
\ea
with $\tau= \sfrac{Q^2}{(M_S+M)^2}$ and
\be
b= e \sqrt{\frac{Q_+^2}{8M (M_S^2-M^2)}}.
\ee

From {\bf equations (\ref{eqF1})-(\ref{eqF2})}
one can make predictions for the form factors and compare 
the obtained results with the experimental data.

\section{Scalar wave functions}
\label{secWF}

Our model is now completely defined, for the baryons 
and for the current, except for  
the scalar function $\psi_{S11}$ which is part of the wave function.

In the spectator quark model the 
wave functions depend
on $(P-k)^2$ only, as the baryon and 
diquark are taken on-mass-shell.
That dependence can be re-written in terms 
of the adimensional variable 
\be
\chi_{_B}= \frac{(M_B-m_D)^2-(P-k)^2}{M_B m_D},
\ee
where $M_B$ is the baryon mass [nucleon or $N(1535)$] 
and $m_D$ the diquark mass.
 
Within the S-wave approach, the  scalar function 
in the nucleon wave function is given by \cite{Nucleon}:
\be
\psi_N(P,k)= \frac{N_0}{m_D (\beta_1+ \chi_{_N}) (\beta_2+ \chi_{_N})},
\label{eqPsiNS}
\ee
where $N_0$ is the normalization constant and 
$\beta_i$ are adimensional parameters 
which measure the momentum scale of the quark-diquark interaction.
As $\beta_2 > \beta_1$, $\beta_2$  
defines the scale for the short distance range
and $\beta_1$ the long distance range.

As the $N(1535)$ corresponds to
a spin 1/2 quark core with the same content 
of the nucleon, it is reasonable 
to consider a form for the scalar wave function
similar to the one taken for the  nucleon
\be
\psi_{S11}(P,k)= \frac{N_1}{m_D (\beta_3+ \chi_{_{S11}}) (\beta_2+ \chi_{_{S11}})},
\ee
where $N_1$ is the normalization constant and 
$\beta_3$ a new range parameter. To start with, 
the same  parameter $\beta_2$ ($\beta_2 > \beta_3$) 
can be used for the two cases, 
the $N(1535)$ and the nucleon,
if one assumes that the two baryons differ 
only in the structure at large distances.
Moreover, on the other hand, and inspired by the
relativistic quark models 
with an harmonic oscillator confinement 
\cite{Capstick95,Aznauryan08}, we consider that 
the nucleon and the $N(1535)$ may as well have 
the same momentum distributions at large distances 
-- as expected for excitations of the same state -- 
and we will thus also take $\beta_3 =\beta_1$.
Then, the nucleon and the $N(1535)$ 
are described by the same 
scalar wave functions in their rest frame.
We may say that this assumption is justified since 
in the chiral limit the nucleon and the $N(1535)$ 
will have the same mass and become 
two different parity states of the same particle.
The difference between the momentum distributions 
in the nucleon and the $N(1535)$ 
come from the difference in 
the orbital angular momentum in their total wave functions.
In the non relativistic limit, this angular dependence 
corresponds to  $Y_{00}(\hat k)$, a constant,
for the nucleon, and $Y_{1m}(\hat k)$, the P-state,  for 
the $N(1535)$.

An alternative parametrization for the scalar 
wave functions would be to force the fit of 
$\beta_3$ to the data and to
introduce a new parameter in our model.
Since we will see that our parameter-free description 
was surprisingly successful, we did not face a good reason to assume
different scalar functions for the nucleon and the $N(1535)$, 
and our results can be considered true predictions,
once the nucleon is correctly described.

\subsection{Overlap integral}

The transition form factors depend on the 
orbital wave functions through their overlap integral ${\cal I}_0$,
defined by Eq.~(\ref{eqInt0}).
Terms that include integrations 
in $k_x$ or $k_y$ 
vanish because of the symmetries 
of the scalar wave function (as function of $\chi_B$),
as shown in Appendix \ref{apCurrent},
and the integral ${\cal I}_0$
carries the signature of the angular 
momentum dependence of the 
nucleon and $N(1535)$ wave functions.

The overlap integral is covariant and it can be 
evaluated in any frame. 
One of the simplest calculations 
is the one that proceeds in the $N(1535)$ (final state) rest frame, 
(see Appendix \ref{apINT}), where
\be
{\cal I}_0  (Q^2)= \int_k \frac{k_z}{|{\bf k}|} 
\psi_{S11} (P_+ \cdot k)\psi_N. (P_- \cdot k) 
\label{eqInt00}
\ee
In the $N(1535)$ rest frame all the angular dependence of 
the wave functions is contained in $\psi_N$, 
given by Eq.~(\ref{eqPsiNS}). This dependence is expressed by 
\be
P_- \cdot k= E E_D + |{\bf q}| k_z, 
\ee 
where $|{\bf q}|$ is the photon three momentum 
in the $N(1535)$ rest frame, as defined in Eq.~(\ref{eqq2}),
$E$ is the nucleon energy, and $E_D$ the diquark energy.
The numerical value of  ${\cal I}_0(0)$
depends therefore on the existing symmetries in the variable $k_z$.
The properties of the overlap  
integral ${\cal I}_0(Q^2)$ are discussed in Appendix \ref{apINT}. 
In particular for small $|{\bf q}|$, one has
\be
{\cal I}_0 (Q^2) \propto |{\bf q}|. 
\label{eqINT2}
\ee
This result has important consequences 
and allows us to define the domain of validity of our model.

In what follows we will label  $|{\bf q}|$ in the 
$Q^2=0$ limit by $|{\bf q}|_0$.
As the photon energy $\omega$ equals
$|{\bf q}|_0$ at $Q^2=0$,
one has then $|{\bf q}|_0 =  \sfrac{M_S^2-M^2}{2M_S}$,
and according to Eq.~(\ref{eqINT2}),   ${\cal I}_0(0) = 0$, if $M_S = M$.
The relation ${\cal I}_0(0)=0$ is then
equivalent to the orthogonality condition
between the $N(1535)$ and the nucleon wave functions.
However, if $M_S\ne M$, the integral  ${\cal I}_0(0)$ 
will be proportional to $|{\bf q}|_0= \sfrac{M_S^2-M^2}{2M_S}$.
Consequently, ${\cal I}_0(0) \ne 0$, and the $N(1535)$ and the nucleon 
wave functions are not exactly orthogonal. 
This result has a dramatic implication since 
the nucleon and the $N(1535)$ should in fact be orthogonal.
This is an artifact  of  the construction of the wave function from 
its non relativistic behavior, and of having imposed to 
it a covariant form with multiplicative scalar functions 
that were not derived from an ab-initio calculation.
A simple picture of what happens is that the nucleon orbital
(S-state) wave function
(defined unambiguously only in the rest frame of the nucleon) 
is distorted by the boost to the rest frame of the $N(1535)$, 
and therefore is not orthogonal
to the $N(1535)$ orbital (P-state) wave function.
This implies that the overlap integral ${\cal I}_0(0)$
does not vanish. 
Still, if the masses of the initial and final state are equal, 
$Q^2=0$ implies $|{\bf q}|_0=0$, as mentioned,
and there is no problem since there is no boost.

The fact that the integral (\ref{eqInt00}) is 
not zero for $Q^2=0$ is therefore a limitation of our 
model when the initial final and initial states have different masses. 
However, the relation (\ref{eqINT2}) can be used 
to establish the range of application of 
the model.
The non orthogonality between the model wave functions of the 
initial and final state decreases  
as $M_S$ approaches $M$. If the mass difference is negligible 
there is orthogonality to a certain extent. 
Then, $|{\bf q}|_0= \sfrac{M_S^2-M^2}{2M_S}$
is a parameter that measures the quality  
of our model approximations to the wave function.
As $|{\bf q}|_0$ corresponds to the photon energy at $Q^2=0$ 
(at the photon point the energy equals the three momentum),
it defines the natural momentum scale of the reaction. 
In the regime $Q^2 \gg |{\bf q}|_0^2$, 
one has ${\cal I}_0(0) \approx 0$, meaning 
that the nucleon and the $N(1535)$ states 
are {\it almost} orthogonal.
As for the physical case $|{\bf q}|_0 \simeq 0.48$ GeV,
${\cal I}_0(0) \simeq 0$ for $Q^2 \gg 0.23$ GeV$^2$, and therefore,
one can say that  $Q^2 > 2.3$ GeV$^2$ establishes the
threshold for the application of our model.

Summarizing, the present model has limitations 
in its applications at low $Q^2$, in particular near $Q^2=0$, 
but can be used in the high $Q^2$ regime, for $Q^ 2  > 2$ GeV$^2$.

\section{Results}
\label{secResults}

\begin{figure}[t]
\centerline{
\mbox{
\includegraphics[width=3.0in]{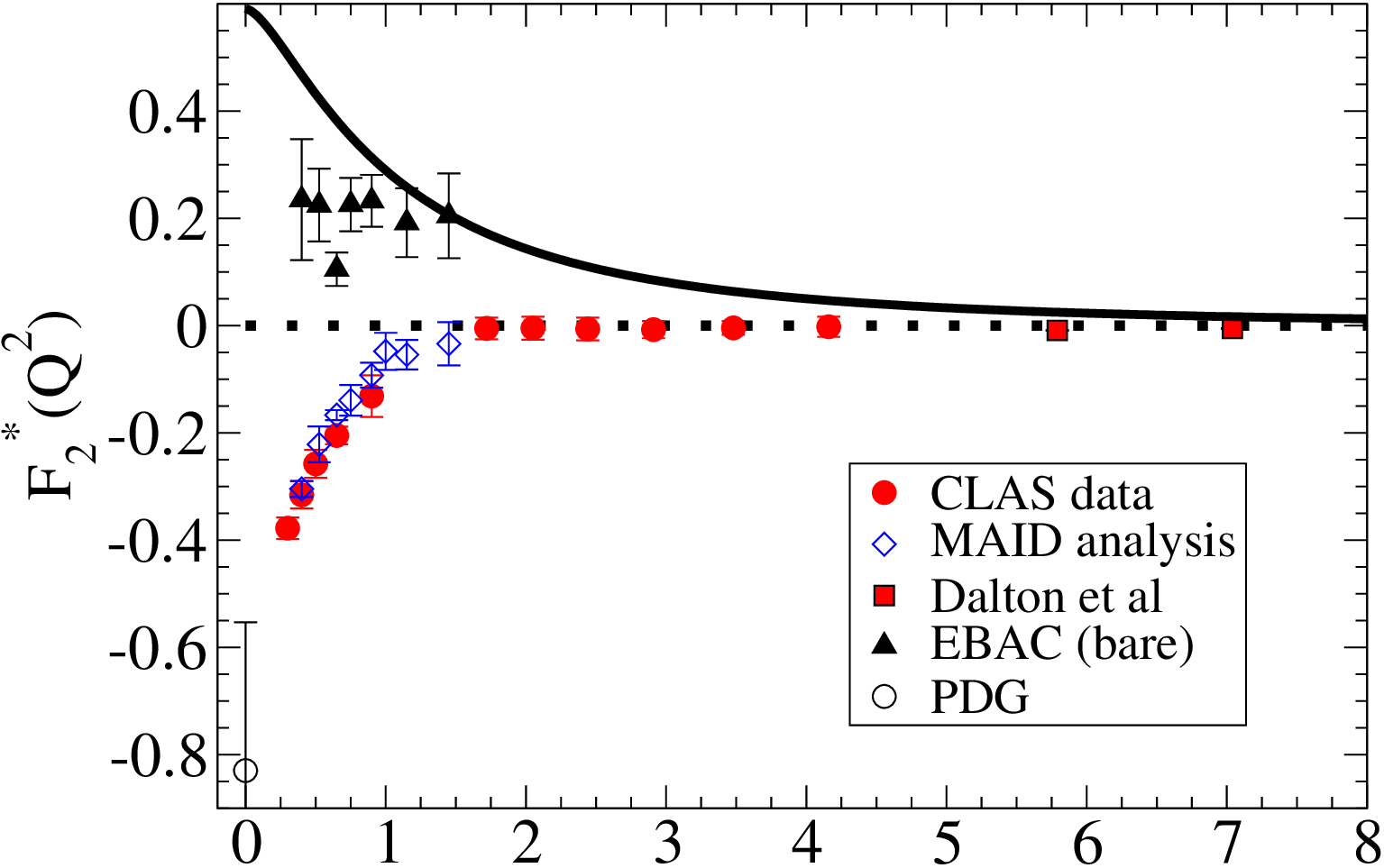} } }
\vspace{.8cm}
\centerline{
\mbox{
\includegraphics[width=3.0in]{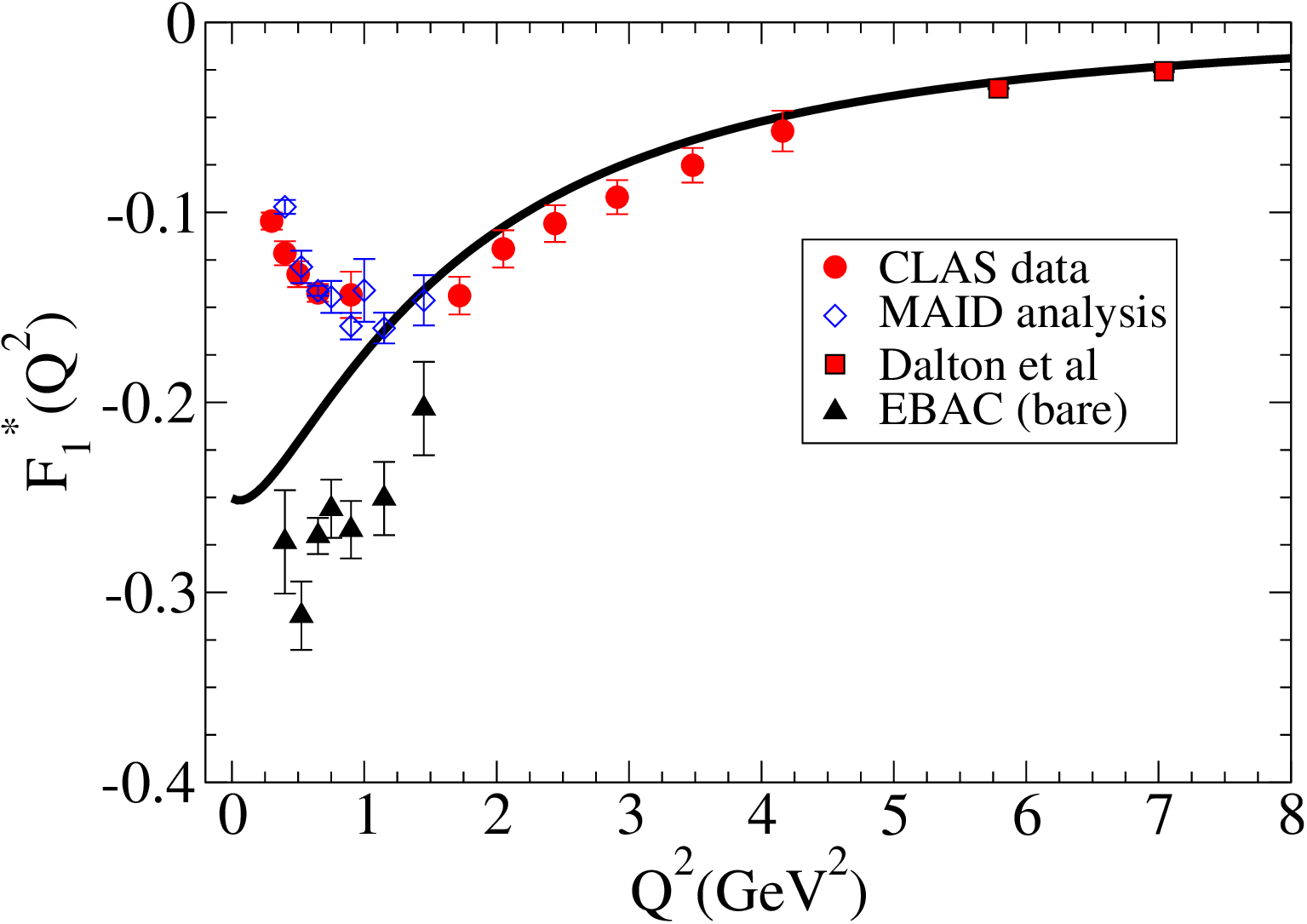} } }
\caption{\footnotesize $\gamma p \to N(1535)$ transition form factors.
CLAS data from \cite{Aznauryan09}, 
MAID data from \cite{MAID}.
The EBAC results \cite{JDiaz09} corresponds to the transition when 
the meson cloud contribution is suppressed.
The solid line is the prediction of the model.
The data for $A_{1/2}(0)$ is given by 
Particle Data Group \cite{PDG}.}
\label{figFF}
\end{figure}

\begin{figure}[t]
\centerline{
\mbox{
\includegraphics[width=3.0in]{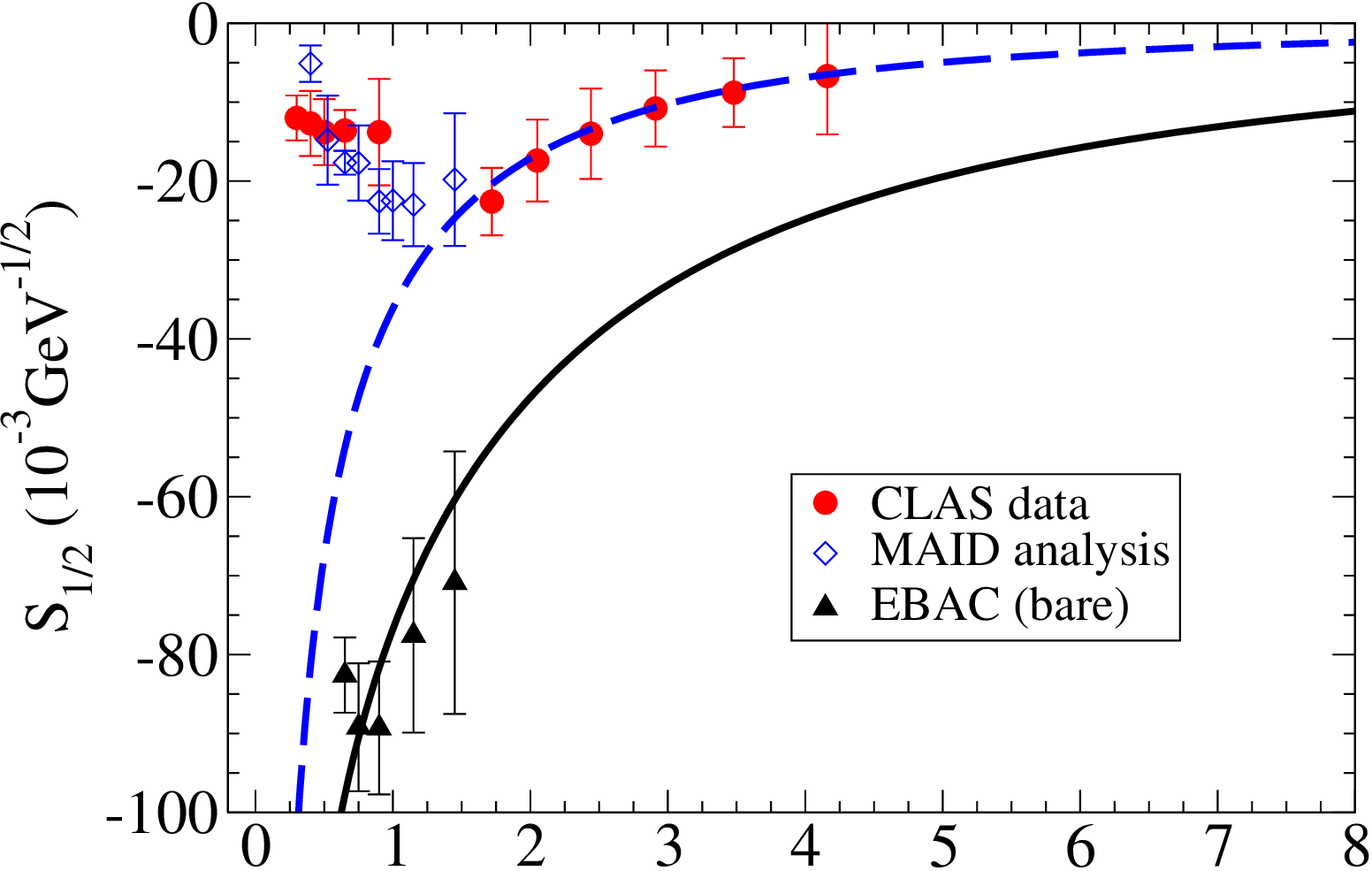} } }
\vspace{.8cm}
\centerline{
\mbox{
\includegraphics[width=3.0in]{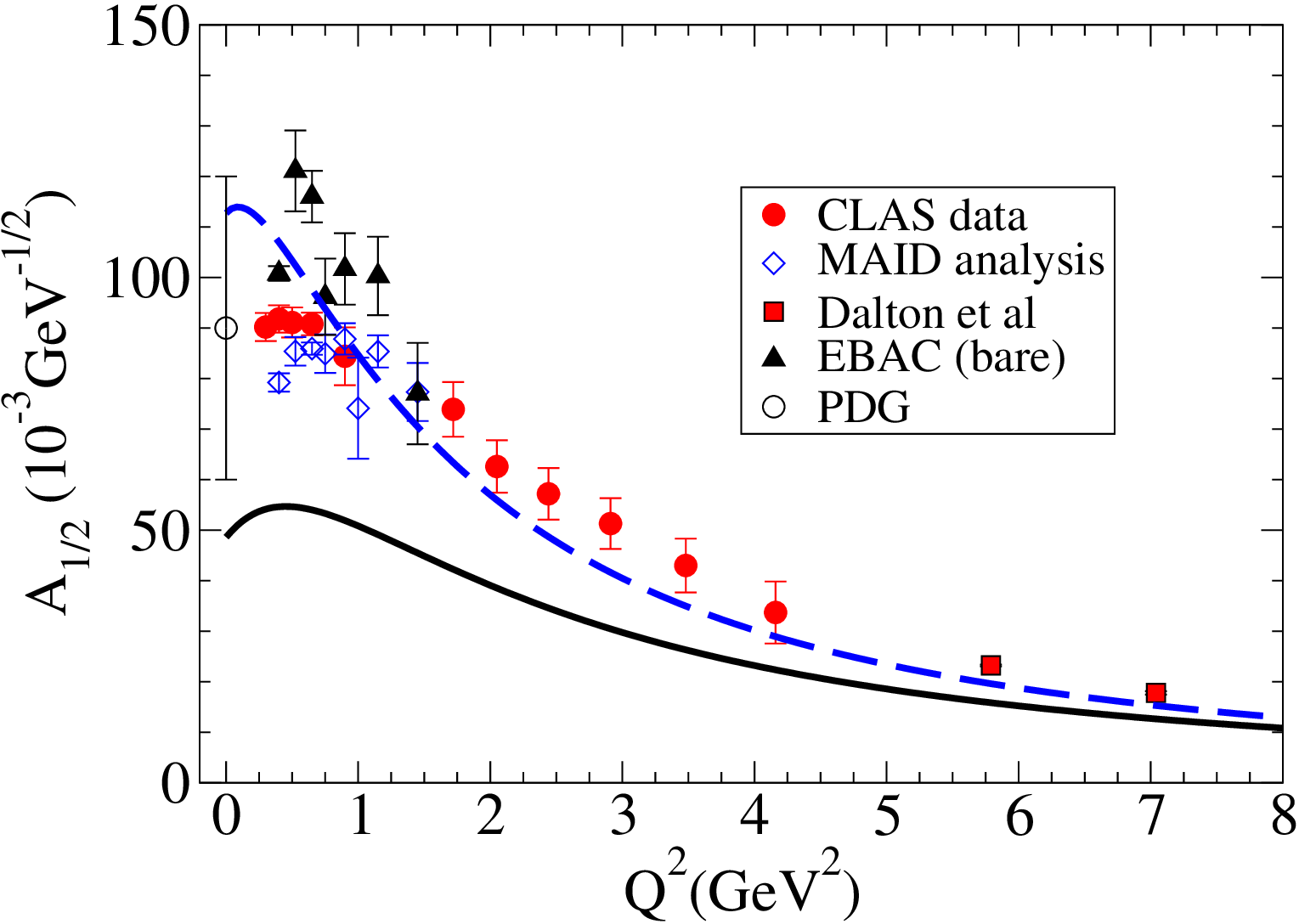} } }
\caption{\footnotesize $\gamma p \to N(1535)$ helicity amplitudes. 
CLAS data from \cite{Aznauryan09}, 
MAID data from \cite{MAID}.
The EBAC results \cite{Aznauryan09} corresponds to the transition when 
the meson cloud contribution is suppressed.
Particle data group data from Ref.~\cite{PDG}.
The solid line is the prediction of the model.
The dashed line is the result under the assumption that $F_2^\ast \equiv 0$
(as supported by the data). }
\label{figAmp}
\end{figure}

With the model for the baryons and for the current depicted in the 
previous sections we have calculates the 
$\gamma N \to N(1535)$ transition 
form factors given by Eqs.~(\ref{eqF1})-(\ref{eqF2})
and the helicity amplitudes given by 
Eqs.~(\ref{eqA12X})-(\ref{eqS12X}).
No parameters of our model were adjusted to these observables.

We calculated only the positive isospin case ($I_z=+1/2$), 
corresponding to the excitation reaction from the proton, 
where the data at finite $Q^2$ for the helicity amplitudes 
is available
\cite{Aznauryan09,MAID,Armstrong99,Thompson01,Denizli07,Dalton09,Brasse84}.
We did not consider the neutron case  ($I_z=-1/2$),
since there is data only for  $Q^2=0$,
and our model is valid only for $Q^2> 2.3$ GeV$^2$. 
The data from  DESY \cite{Brasse84}
and  from Jefferson Lab \cite{Armstrong99,Thompson01,Denizli07,Dalton09}
are restricted only to the $A_{1/2}$ amplitude, 
assuming that the amplitude 
$S_{1/2}$ was negligible.
That assumption was contradicted by the recent 
CLAS \cite{Aznauryan09} and MAID \cite{MAID} analysis.
In the following we use 
Ref.~\cite{Aznauryan09,MAID} 
where $A_{1/2}$ and $S_{1/2}$ were determined simultaneously.
We will also compare our results with the Dalton \etal $\,$ data
\cite{Dalton09},
for $A_{1/2}$ at high $Q^2$ ($Q^2> 5.4$ GeV$^2$), which is
determined under the assumption that $S_{1/2}=0$
[for large $Q^2$ 
the approximation $S_{1/2}=0$
is better justified  due to the 
falloff of $S_{1/2}$ at high $Q^2$].

\subsection{Transition form factors}

The results for the 
$\gamma N \to N(1535)$ form factors 
are shown in Fig.~\ref{figFF}.
The data for $F_1^\ast$
and $F_2^\ast$ was obtained by inverting the relations 
(\ref{eqA12X})-(\ref{eqS12X}).
In the figure we represent also 
the CLAS data from Ref.~\cite{Aznauryan09}
and the MAID analysis of Ref.~\cite{MAID}, 
as well as the results from~\cite{Dalton09} (where $S_{1/2}=0$).
One can 
see that our model  
describes well the $F_1^\ast$ data for  
for $Q^2 > 1.5 $ GeV$^2$, in particular 
that the model works in its regime 
of application  $Q^2 > 2.3 $ GeV$^2$.
As for $F_2^\ast$, our model fails completely
when compared with the experimental data.
We predict positive values for $F_2^\ast$,
contrarily to the data.
Also, the magnitude differs strikingly from 
the data: the CLAS data is very close to zero
for $Q^2 > 2$ GeV$^2$, in the region 
where our model gives a strong positive contribution.
This disagreement can be interpreted in two ways.
One possibility is that our model is limited 
because the internal diquark P-states 
were neglected in our model, and we will have to confirm 
their effects in a future work.
Other possible interpretation
is that, for $F_2^\ast$ the valence quark effects,
the only ones considered in our model
are  strongly canceled by
the effect of the meson cloud polarization, not included in our model.
If this last interpretation is correct, 
one has to conclude that  meson cloud effects 
are very significant, even in the 
region $Q^2 > 2$ GeV$^2$. This finding  is at odds
with what was observed till now in 
similar systems, like the nucleon \cite{Nucleon}
and the Roper \cite{Roper}. Nevertheless,  
the $\gamma N \to \Delta$ quadrupole form 
factors reveal a  strong contribution  of strong pion cloud 
in the region 2--6 GeV$^2$ 
\cite{NDeltaD,LatticeD}.

To test the last interpretation, we compared 
our valence quark model predictions with the
calculations from a different framework,
the EBAC dynamical coupled-channel model 
based in Sato-Lee model~\cite{Matsuyama07}.
In the EBAC analysis~\cite{JDiaz09} the 
effects of the meson cloud dressing are subtracted, 
and the pure quark core contributions calculated 
from the model. 
The EBAC data can then be directly compared with our results,
as shown also in Fig.~\ref{figFF} (upper triangles).
As for $F_1^\ast$,  the EBAC results 
overestimates (in absolute value) 
the experimental data (CLAS and MAID) 
but seems to approach the data for $Q^2 \approx 2$ GeV$^2$.
As for $F_2^\ast$,  the EBAC results are surprisingly
consistent with our own predictions, both in sign and magnitude
for $Q^2 \approx 1.5$ GeV$^2$,
near the threshold where our model 
starts to be applicable, $Q^2 > 2.3$ GeV$^2$. 
Future EBAC determination of the quark core contributions, 
already planed for higher $Q^2$ \cite{Kamano10}, 
will be very important to test our predictions and 
interpretations.
An independent confirmation of the 
large contribution of the valence quarks 
for $F_2^\ast$ may also come from lattice QCD 
at high $Q^2$. We note that our covariant spectator quark model 
was already successful in the description 
of lattice QCD simulations for the nucleon, 
Roper \cite{Nucleon,Roper,ExclusiveR} 
and $\Delta$ systems \cite{Lattice,LatticeD}. 

To summarize, our results for the form factor $F_1^\ast$ 
are consistent with the data for $Q^2 > 2$ GeV$^2$, 
in the domain of validity of our model.
As for $F_2^\ast$, our model supports the 
idea that meson cloud contributions are comparable 
with the valence quark contributions, 
which is also validated by the EBAC studies 
of the $N(1535)$ system \cite{JDiaz09}.

\subsection{Helicity amplitudes}

Using our results for the form factors 
we have also calculated the helicity amplitudes 
in the $N(1535)$ rest frame, corresponding 
to the transformations (\ref{eqA12X})-(\ref{eqS12X}).
Some comments are necessary before 
showing the results.
The first note is that our quark model should be compared 
with the data only in the region $Q^2 > 2.3 $ GeV$^2$.
A second important note is that in our model 
$F_1^\ast(0) \ne 0$ because of the violation of 
the orthogonality condition between the 
nucleon and the $N(1535)$ wave functions. 
Therefore the amplitude $S_{1/2}$ in our model 
is singular for $Q^2=0$, 
in opposition to the finite result
expected from the data. 
This effect was already reported 
in the relativistic quark model of Ref.~\cite{Konen90},
where the quark current was modified to 
restore gauge-invariance.
With those limitations in mind, 
we represent in Fig.~\ref{figAmp} 
the amplitudes corresponding to the form factors
in Fig.~\ref{figFF}, by the solid line.
The dramatic deviation from the data 
is not surprising, since our model 
disagrees already with the $F_2^\ast$ data. 
The disagreement is evident for $A_{1/2}$ 
where our large $F_2^\ast$ contribution 
spoils a excellent result that would be obtained 
if the $F_2^\ast$ could be  neglected. 
The results obtained in that scenario ($F_2^\ast(Q^2) \equiv0$) 
are represented by the dashed line.
In that case the agreement of our model 
with the data is excellent for $Q^2 > 2$ GeV$^2$
for both amplitudes.
It is moreover interesting to 
note that the model (solid line) agrees well 
with the EBAC results for $S_{1/2}$. That 
comes from the $F_1^\ast$ suppression 
in the $S_{1/2}$ amplitude 
by the factor $\sfrac{M_S-M}{M_S+M}$
[see Eq.~(\ref{eqS12X})].

We conclude that 
the helicity amplitudes are not the best representation
to test our model, since those amplitudes amplify
the limitations of our model, like $F_1^\ast(0) \ne 0$ 
or the large magnitude of $F_2^\ast$.
Combining our results for $F_1^\ast $with the assumption 
that $F_2^\ast$ is negligible for $Q^2 > 2$ GeV$^2$,
as a consequence of the meson cloud effect, which is 
substantiated by the data  and the EBAC results,   
one can achieve a very good description of the 
helicity amplitudes data.

\subsection{Comparison with the literature}
\label{secDiscussion}

The study of the $\gamma N \to N(1535)$ electromagnetic 
structure was in the past based almost only on  
the representation of the helicity amplitudes 
[in the $N(1535)$ rest frame].
Then, the comparison with other works has to be done in this representation.
From the previous section we know that 
the data corresponds to positive values for 
$A_{1/2}$
and negative values for $S_{1/2}$.  

We will start by discussing the constituent quark models. 
Different quark model predictions, including
non-relativistic \cite{Warns90,Close90,Aiello98,Aznauryan08,Zhao02} 
and relativistic \cite{Capstick95,Konen90,Pace,He06} formulations,
agree qualitatively with the data for $A_{1/2}$.
In particular, in Ref.~\cite{Capstick95},  calculations based
on the light-front formalism give an 
excellent description of the $A_{1/2}$ data for $Q^2 > 2$ GeV$^2$
\cite{Aznauryan09}.
Also QCD sum rules \cite{Braun09} 
are consistent with the $A_{1/2}$ data for $Q^2 > 1$ GeV$^2$.

In a non relativistic model 
with harmonic-oscillator confinement potential 
the relative 
sign between $A_{1/2}(0)$ and $S_{1/2}(0)$ 
is positive, and determined by the relative sign between 
the $\pi N N$ and the $\pi N N(1535)$ coupling constants
\cite{Aznauryan08}. 
For non relativistic models we should expect 
then  positive values for $S_{1/2}$ at low $Q^2$.
This 
feature is also shared by light-front and relativistic quark models 
\cite{Capstick95,Konen90,Warns90,Pace,Aznauryan09} 
although sometimes
negative results are obtained for $Q^2 > 2$ GeV$^2$
\cite{Capstick95,Pace,Aznauryan09}.
Still, in general one has  the same
sign for $A_{1/2}(0)$ and $S_{1/2}(0)$.
Exceptions to this feature are obtained by 
the QCD sum rules \cite{Braun09} and our model.
QCD sum rules predict the sign but underestimate
in absolute value the result for $S_{1/2}$.

It has also been suggested that the state $N(1535)$
may have a strong contribution from quark-antiquark states,
or even been dynamically generated by 
the meson-baryon interaction.
An and Zou~\cite{An08} considered 
a quark model with 
explicit quark-antiquark dressing, and
concluded that those effects can be of the order 
of 20\% for low $Q^2$.
In the overall, the signs and magnitudes are consistent with the data.
In Ref.~\cite{Golli11} the meson cloud dressing 
is calculated within the cloudy bag model. 
In that case the quark core is dominant at low $Q^2$ 
and is consistent with the data for $A_{1/2}$ 
(with $\approx 25\%$ of meson cloud), although 
the $S_{1/2}$ data is overestimated.
For $Q^2> 1.5$ the model predictions are 
suppressed compared with the data  
indicating that short range behavior 
is not well simulated by the bag model~\cite{Golli11}.

The helicity amplitudes were also determined 
using a chiral unitary approach \cite{Jido08,Hyodo08}.
The authors conclude that the $N(1535)$ 
seems to be largely dynamically generated 
from the interaction of mesons and baryons 
but also that a genuine quark component 
is necessary particularly at high $Q^2$ \cite{Jido08}.
Qualitatively, the meson dressing explains roughly
50-60\% of the $A_{1/2}$ amplitude.
Also, the calculations of the EBAC group sugest
the importance of the meson dressing 
at low $Q^2$, 
although there is a dominance of the quark core \cite{JDiaz09}.

One may conclude that in general, from quark 
models and hadronic models with meson dressing, 
the  meson cloud can be important, 
but genuine valence quark contributions 
are equally necessary to explain the data.

\subsection{Large $Q^2$ regime}

\begin{figure}[t]
\centerline{
\mbox{
\includegraphics[width=3.0in]{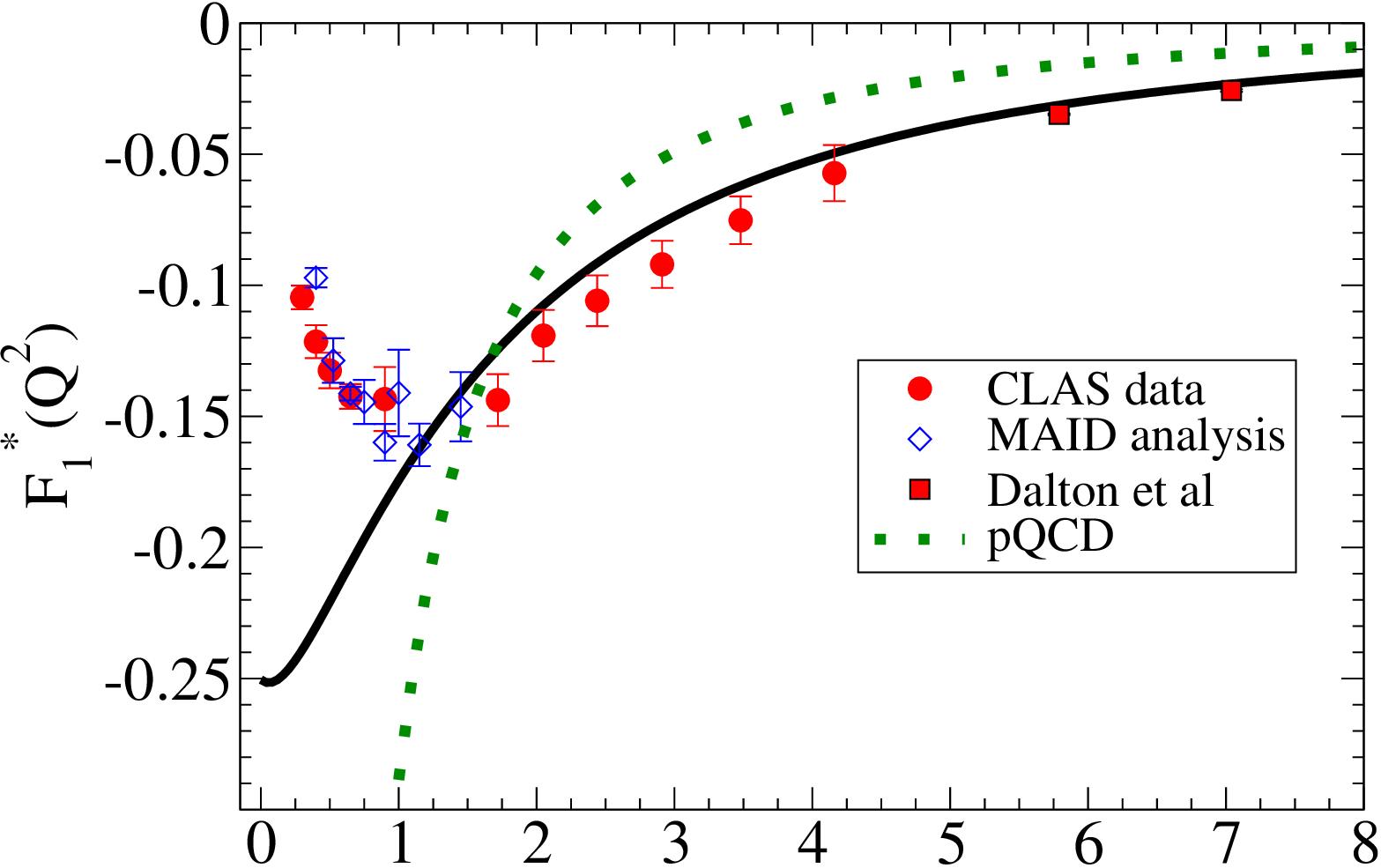} } }
\vspace{.8cm}
\centerline{
\mbox{
\includegraphics[width=3.0in]{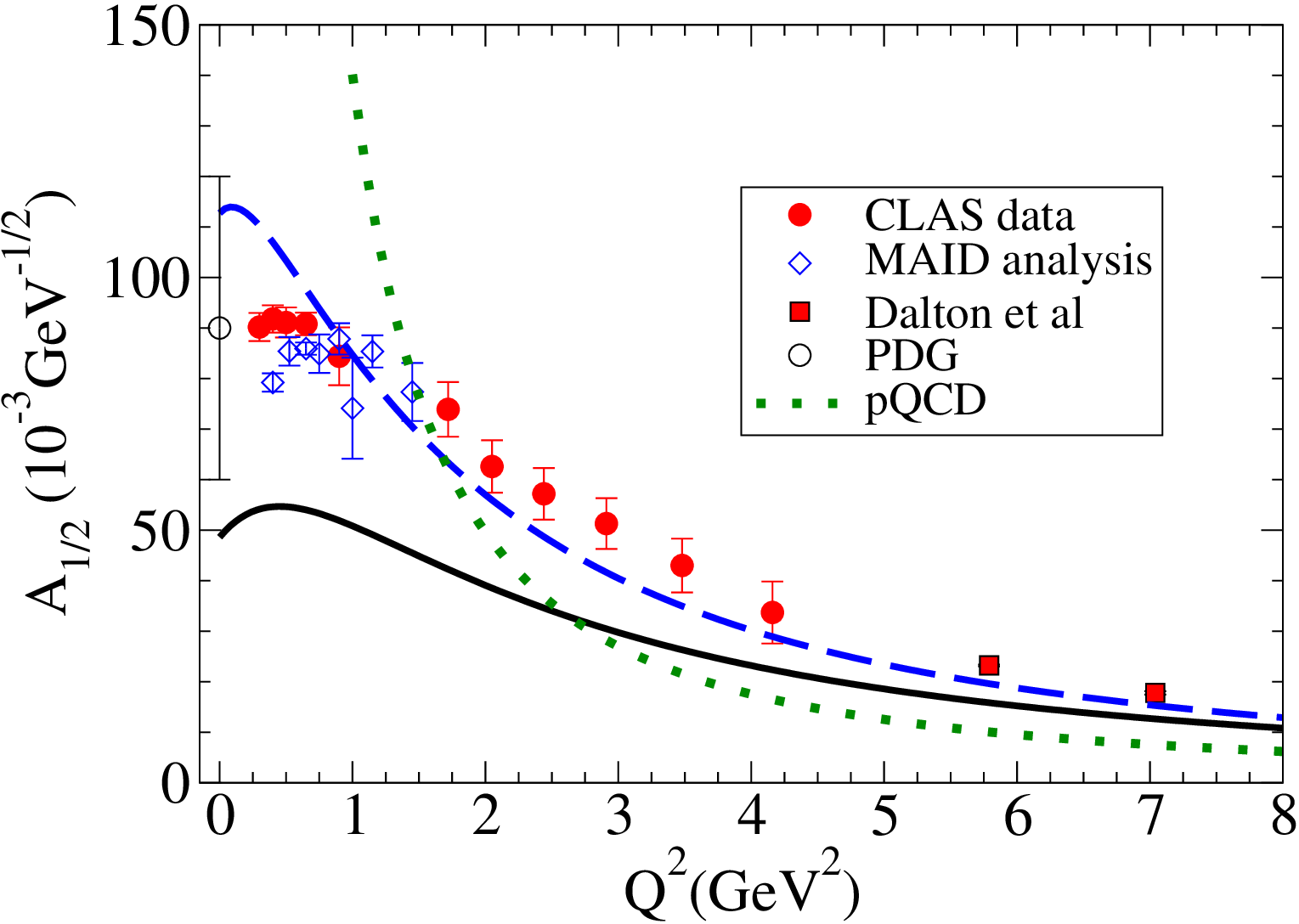} } }
\caption{\footnotesize 
$\gamma p \to N(1535)$ transition.
Comparing $A_{1/2}$ amplitude and 
$F_1^\ast$ with the asymptotic expressions.
Same meaning of previous figures. 
The pQCD result is from Ref.~\cite{Carlson}.}
\label{figASSYMP1}
\end{figure}

The study of the asymptotic dependence of the 
$\gamma N \to N(1535)$ 
transition form factors attracts some attention,
because pQCD predicts a very slow falloff 
for $A_{1/2}$~\cite{Carlson} and also 
because precise experimental data have been extracted 
at high $Q^2$,  in particular for 
$Q^2 \approx 4$ GeV$^2$~\cite{Armstrong99}
and $Q^2 \simeq 5.7$, and $7.3$ GeV$^2$~\cite{Dalton09}.
The estimate from pQCD~\cite{Carlson} 
for large $Q^2$  is 
\be
Q^3 A_{1/2}(Q^2) = e \sqrt{\frac{M}{M_S^2-M^2}} \beta,
\label{eqA12assymp}
\ee
where $\beta= 0.58$ GeV$^3$, in the more optimistic 
estimate (upper limit)~\cite{Carlson}.
As for the form factors, one expects 
$F_1^\ast \sim \sfrac{1}{Q^4}$ and  $F_2^\ast \sim \sfrac{1}{Q^6}$, 
apart $\log Q^2$ corrections. 
Then, for large $Q^2$, one has $|F_1^\ast| \gg |F_2^\ast|$, 
and according  to Eq.~(\ref{eqA12X})
\ba
Q^4 F_1^\ast(Q^2) = 
- \sqrt{\frac{2 M^2 Q^2}{(M_S+M)^2 + Q^2}} \beta.
\label{eqF1assymp}
\ea
The asymptotic results from  Eqs.~(\ref{eqA12assymp})-(\ref{eqF1assymp}) 
are presented in Fig.~\ref{figASSYMP1}
for both $F_1^\ast$ and $A_{1/2}$.
In the last case we show the result  obtained by making $F_2^\ast = 0$, 
as discussed earlier.
In the figure it is clear that the pQCD 
estimation underestimates the data and our model 
for high $Q^2$.

\begin{figure}[t]
\centerline{
\mbox{
\includegraphics[width=3.2in]{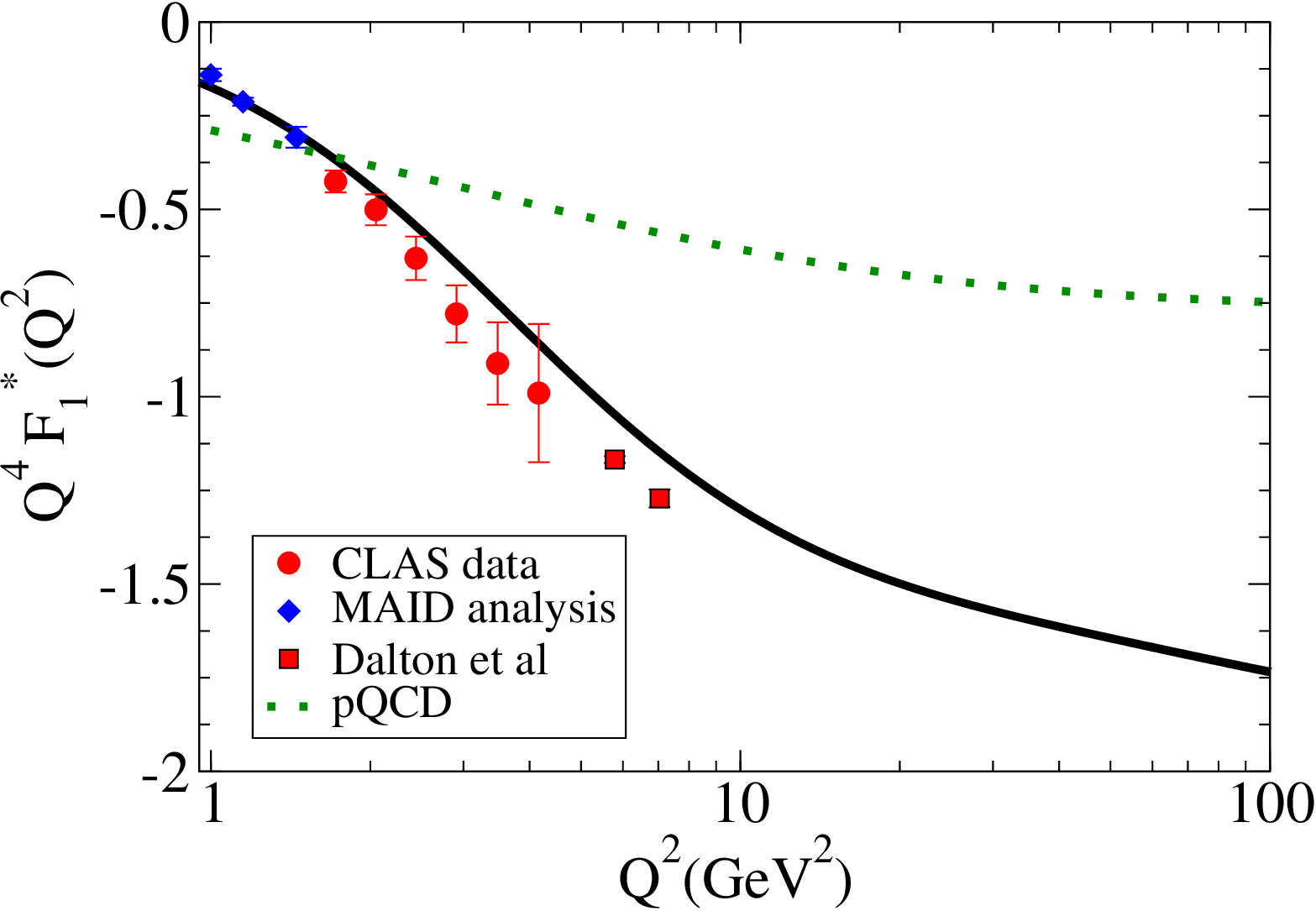} } }
\caption{\footnotesize 
$Q^4F_1^\ast(Q^2)$ for high $Q^2$ compared with the data.  
The model (solid line) 
can be represented for high $Q^2$ 
as $Q^4 F_1^\ast (Q^2) \simeq - 0.144\log \frac{Q^2}{\Lambda^2}$
with  $\Lambda^2= 0.4982$ GeV$^2$.}
\label{figF1as}
\end{figure}

The asymptotic behavior of the form
factors can be better understood scaling
the functions by a convenient power of $Q^2$
to check if the results converge to a constant,
apart the logarithm corrections.
In this case we should take the functions 
$Q^3 A_{1/2}$ and $Q^4 F_1^\ast$.
The results for $F_1^\ast$ are presented in Fig.~\ref{figF1as}.
The representation of $A_{1/2}$ would be equivalent.
In the figure it is clear that pQCD estimation 
fails the description of the data 
by a factor larger than 2.
The same  was reported in Ref.~\cite{Dalton09} for $A_{1/2}$.
The pQCD prediction differs then  
from the spectator quark model.
At $Q^2=100$ GeV$^2$ the ratio is 2.3.
Also in the figure it is clear a non constant slope for both pQCD 
and the spectator quark model results in the region shown, 
indicating corrections for the $1/Q^4$ behavior.
In the pQCD case, the slope is a consequence of the 
$Q^2$-dependent factor of the r.h.s.~of 
Eq.~(\ref{eqF1assymp}), which became a constant 
only for $Q^2 \gg (M_S+M)^2 = 6.1 $ GeV$^2$ 
[see the slow variation of the dotted line in Fig.~\ref{figF1as}].
As for the spectator quark model, the logarithm 
dependence at larger $Q^2$, comes from the 
parametrization of the nucleon wave functions
by Eq.~(\ref{eqPsiNS}), as product 
of two monopole factors in the variable $(P-k)^2$.
That choice was considered in the applications 
to the nucleon electromagnetic structure 
\cite{Nucleon} in order to reproduce the 
expected pQCD behavior for 
the nucleon form factors 
(Dirac $F_1 \sim \frac{1}{Q^4}$ and Pauli 
$F_1 \sim \frac{1}{Q^6}$), but also 
contains logarithm corrections.
See Appendix G from Ref.~\cite{NDelta} for details.

For $Q^2 > 20$ GeV$^2$ one can represent 
the spectator quark model form factor $F_1^\ast$ as  
\be
F_1^\ast (Q^2) \simeq - \frac{0.144}{Q^4}\log \frac{Q^2}{\Lambda^2},
\ee
where $\Lambda^2= 0.4982$ GeV$^2$.

In conclusion, our model reveals a scaling 
with the same power as pQCD for $Q^2 \approx 100$ GeV$^2$, 
apart logarithm corrections.
The scaling due to pQCD, if it is confirmed, 
will be revealed only for much larger $Q^2$ values than in our model.

\section{Conclusions}
\label{secConclusions}

In this work we applied the covariant 
spectator quark model to the $N(1535)$ system.
We considered the simplest case 
where  $N(1535)$  is made of
states with core spin 1/2, and we
neglected the effect of the  core spin 3/2 state,
as in Ref.~\cite{Aznauryan08}.
We took also the diquark as a pointlike particle
(no internal P-states).
These approximations have the advantage of 
reducing the degrees of freedom of our model to the minimum,
and to allow us to perform calculations with no adjustable parameters,  
since all parameters 
(in the quark current and wave functions)
were already fixed 
by the study of the nucleon system \cite{Nucleon}.
Our results in this paper are then true predictions.
The extension of this work  
to include spin 3/2 cores (which are also part of the nucleon D-states)
is in progress \cite{InProgress}.
Once our model is calibrated for 
the spin 3/2 component we will also be able of 
making predictions for the $N(1650)$ form factors.

Our model takes  
contributions for the form factors from the 
valence quarks  alone, and
neglects possible meson cloud effects 
(in principle dominated by $\eta$ and $\pi$ clouds). 
This approximation involving the meson cloud 
suppression simplifies the construction 
of the $N(1535)$ wave function 
(as a three-quark system). 
Another approximation, intrinsic to the relativistic generalization 
that we make for the wave function, is that 
the $N(1535)$ state is exactly orthogonal to 
the nucleon state only in the case of equal masses for the two baryons $M_S=M$.
However, as the orthogonal condition 
can be written in powers of $(M_S-M)$, 
one can show that our results are accurate 
for $Q^2 > 2.3$ GeV$^2$.
In that region meson cloud effects are 
expected to be negligible, the reason why 
one can make predictions for the 
form factors, which otherwise would contain, apart from valence quark effects, 
important meson cloud contributions.
    
For the $F_1^\ast$ form factor our results are in excellent 
agreement with the data in  the domain
of applicability of our model.
This is remarkable since there is no parameter adjustment.
Our results for $F_1^\ast$ are also close 
to the EBAC analysis of the quark core effects, 
although the EBAC results are restricted 
to the region $Q^2 < 2$ GeV$^2$.
  
As for the $F_2^\ast$ form factor, our predictions fail completely 
to describe the experimental data in 
their sign and magnitude,  which is consistent with 
$F_2^\ast \simeq 0$ for  $Q^2 > 2$ GeV$^2$.
Our results are however in good agreement with 
the estimations of the EBAC group 
of the quark core contribution to the $F_2^\ast$ form factor
near $Q^2=2$ GeV$^2$.
These two last points suggest that our failure in describing  $F_2^\ast$ 
is caused by a large negative contribution from the meson cloud 
which  cancels almost exactly the valence quark contribution. 
Although meson cloud contributions 
are expected to decrease with increasing $Q^2$, 
there are some exceptions to that rule, 
as the observed for the $\gamma N \to \Delta$ 
quadrupole transition form factors 
\cite{NDeltaD,LatticeD}, where pion cloud 
are in fact the dominant effect.
The other  possible explanation for the
failure of our model in the description of the $F_2^\ast$, 
is the internal structure of the diquark which was not considered here. 
But this explanation  is excluded by 
the comparison of our results with the EBAC result,  
which seems to indicate that the  pointlike diquark
approximation is apparently good, at least for $F_1^\ast$.

A true test of the $F_2^\ast$ suppression 
can come from the extraction of the 
core contributions by EBAC model 
for higher $Q^2$, planned for a near future \cite{Kamano10}, and
which can confirm our  results for the  
valence quark contributions.
That test will also be useful to assert and consolidate our $F_1^\ast$ results.
A third independent test can be the 
direct comparison with lattice QCD simulations, 
particularly for large pion masses (say $m_\pi > 0.4$ GeV), 
a regime where quark-antiquark ($\pi$ and $\eta$ cloud) 
contributions are believed to be very small.
Lattice QCD simulations are nowadays viable
since they were performed previously 
for the $\gamma N \to \Delta$ 
and $\gamma N \to N(1440)$ reactions \cite{Alexandrou08,Lin09}.
Although the comparison of phenomenological model results, 
at the physical pion mass point,
with lattice QCD can be problematic 
due to the necessity of extrapolating 
to the physical limit, that is not a problem 
for the spectator quark model: it is based on a vector meson dominance 
parametrization of the current, and therefore can be extended 
successfully to the lattice conditions, 
as was shown for the nucleon \cite{Lattice} and  
Roper \cite{Roper} reactions, 
for the  $\gamma N \to \Delta$ transition
\cite{Lattice,LatticeD}  
and also for the baryon decuplet form factors \cite{Omega}.

In addition to the form factors, we calculated as well the 
helicity amplitudes $A_{1/2}$ and $S_{1/2}$.
As in our calculations 
the  violation of the orthogonality condition between 
the initial and final states, gives  
$F_1^\ast(0) \,  \propto \, {\cal I}_0(0) \ne 0$,
implying that the amplitude
$S_{1/2}$ diverges for $Q^2 \to 0$
and the results for $F_2^\ast$ 
differ from the data, 
we conclude that helicity amplitudes 
are not the more convenient representation  
to test our model in particular,
and quark models in general. 
Combining our results with the hypothesis 
that $F_2^\ast$ is negligible,
because of the actual cancellation 
of valence quark contributions and meson cloud contributions,
which is suggested by the successful comparison of 
the our results and the EBAC quark core contribution, 
we obtain an excellent description 
of the helicity amplitudes data, $A_{1/2}$ and $S_{1/2}$
(see dashed line in Fig.~\ref{figAmp}).
As for $A_{1/2}$ the agreement is remarkable 
for $Q^2 > 1$ GeV$^2$, even before the region 
of validity of our model  is reached.
As for  $S_{1/2}$, although it is singular for $Q^2=0$, 
the model describes  the data 
for $Q^2 > 1.5$ GeV$^2$. 
 
In summary, the $\gamma N \to N(1535)$ reaction is  
very interesting from the constituent quark model perspective.
The possibility of the $F_2^\ast$ form factor to vanish at  
intermediate $Q^2$ values, 
in contrast to what happens  with all other known resonances, 
provides a unique challenge 
to theoretical models, in order to understand 
the role of the valence quarks, and their interplay with
the meson cloud. 
All effort from quarks models, 
dynamical coupled-channel reaction models, 
chiral effective models and lattice QCD, 
are welcome in attempts that have to be harmonized and supplemented together, 
in order to interpret the 
$\gamma N \to N(1535)$ reaction data.


\vspace{0.3cm}
\noindent
{\bf Acknowledgments:}

\vspace{0.2cm}

The authors thank Hiroyuki Kamano for providing the 
EBAC results from Ref.~\cite{JDiaz09} 
and Viktor Mokeev for helpful discussions. 
The authors also thank Franz Gross 
for the invitation to visit 
the Jefferson Lab Theory Group.
G.~R.\ was  supported by the Funda\c{c}\~ao para
a Ci\^encia e a Tecnologia under the Grant
No.~SFRH/BPD/26886/2006.
This work is also supported partially by the European Union
(HadronPhysics2 project ``Study of strongly interacting matter'').
The work of the two authors was also financed by 
the Funda\c{c}\~ao para a Ci\^encia e a Tecnologia,
under grant No.~PTDC/FIS/113940/2009, 
``Hadron structure with relativistic models''.

\appendix

\section{$N(1535)$ wave function}
\label{apWF}

\subsection{Non relativistic form}

The $N(1535)$ is defined as an excitation of the nucleon 
corresponding to the state $I(J^P)=\sfrac{1}{2}\left(\sfrac{1}{2}\right)^-$.
To represent the  $N(1535)$ state in a constituent 
quark model framework we need  
to consider the momentum, spin, isospin of each quark
and relate it with the $N(1535)$ proprieties.
We will follow the construction based 
on the $SU(6)\otimes O(3)$ 
as in Refs.~\cite{Giannini91,Capstick00,Isgur77,Isgur79,Warns90,Close90,Li90}.

\subsubsection{Jacobi momenta}

We label the momentum of quark $i$ by $k_i$.
The center of mass momentum $P$ is then 
given by $P=k_1+ k_2+ k_3$.
At this point we do not distinguish 
between non relativistic and relativistic kinematics.
The Jacobi momentum 
are
\be
k_\rho= \frac{1}{\sqrt{2}}(k_1-k_2), 
\ee
for the relative momentum of the quark 
in the quark-pair (12),  and 
\be
k_\lambda=
\frac{1}{\sqrt{6}}(k_1+k_2-2k_3), 
\ee
to measure the relative momentum
between the diquark center of mass and 
the third quark.
Note that $k_\rho$ is anti-symmetric 
in the exchange of quarks 1 and 2,
and that $k_\lambda$ remains unchanged 
(symmetric) in the same exchange.

We note that 
in the non relativistic limit and  
in the baryon center of mass frame
(${\bf k_1}+{\bf k_2}+ {\bf k_3}={\bf 0}$) one has 
${\bf k}_3=-({\bf k}_1+{\bf k}_2)$. 
Therefore,
\be
{\bf k}_\lambda=  \sqrt{\frac{3}{2}} {\bf k},
\label{eqKlambda}
\ee
where ${\bf k}={\bf k}_1+{\bf k}_2$ is the diquark three momentum.

We will use the $\rho$ and $\lambda$ labels to 
characterize the baryon states, as it was defined in the main text, 
and  as it is usual practice in the
literature, e.g.~in Ref.~\cite{Capstick95,Aznauryan08}.

\subsubsection{Spin states}

In the coupling of the spins of the 3 quarks 
there are different combinations for $(s_{12},s)= (|{\bf s_1 + s_2}|,s)$,
where $s_{12}$ is the sum of  the spins of quarks (12) 
and $s$ the spin of quark (3).
The possible combinations are
\ba
\chi^\rho =\left(0,\frac{1}{2}\right), \hspace{.3cm}
\chi^\lambda= \left(1,\frac{1}{2}\right),  \hspace{.3cm}
\chi^S= \left(1,\frac{3}{2}\right), 
\nonumber 
\ea 
respectively the $\rho$-type  ($\chi^\rho$) and the 
$\lambda$-type ($\chi^\lambda$)  states with mixed symmetry, 
and the state ($\chi^S$) which is symmetric 
the change of any of the three quarks.

The spin states $\chi^\rho$ and $\chi^\lambda$
are defined in terms of combinations of two spin states 
[quark pair (12)],
anti-symmetric and symmetric respectively,
with the spin of the quark 3.
This construction is  
similar to what was done for the nucleon \cite{Nucleon,NDelta}.
One has for the spin projection $+ 1/2$:    
\ba
\chi^\rho (+\sfrac{1}{2}) &\equiv&
\ket{\sfrac{1}{2}, +\sfrac{1}{2}}_\rho \nonumber \\
& = &
\frac{1}{\sqrt{2}} \left( \uparrow \downarrow- 
\downarrow \uparrow   \right) \uparrow \\
\chi^\lambda (+\sfrac{1}{2}) &\equiv&
\ket{\sfrac{1}{2}, + \sfrac{1}{2}}_\lambda 
\nonumber \\ 
& =& \frac{1}{\sqrt{6}} \left( 2 \uparrow \uparrow \downarrow- 
\uparrow \downarrow \uparrow  -
\downarrow \uparrow \uparrow \right).
\ea
Identical expression hold for the isospin states.
For example, for the proton  (isospin projection $+1/2$),  
we write the isospin states as 
\ba
& &
\phi_I^0  (+\sfrac{1}{2}) \equiv 
\frac{1}{\sqrt{2}}(ud-du)u \nonumber \\
& &
 \phi_I^1  (+\sfrac{1}{2}) \equiv 
\frac{1}{\sqrt{6}}(2uud-udu-duu),
\ea 
preserving the notation used in the nucleon 
wave function \cite{Nucleon}.
Here the anti-symmetric state in the pair  is identified by 0 and 
the symmetric state by 1.

For completeness, we represent also the state corresponding
to isospin and spin projections $-1/2$:
\ba
& &
\phi_I^0  (-\sfrac{1}{2}) \equiv 
\frac{1}{\sqrt{2}}(ud-du)d \nonumber \\
& &
 \phi_I^1  (-\sfrac{1}{2}) \equiv 
-\frac{1}{\sqrt{6}}(2ddu-udd-dud),
\ea 
\ba
\chi^\rho (-\sfrac{1}{2}) &\equiv&
\ket{\sfrac{1}{2}, -\sfrac{1}{2}}_\rho \nonumber \\
& = &
\frac{1}{\sqrt{2}} \left( \uparrow \downarrow- 
\downarrow \uparrow   \right) \downarrow \\
\chi^\lambda (-\sfrac{1}{2}) &\equiv&
\ket{\sfrac{1}{2}, - \sfrac{1}{2}}_\lambda 
\nonumber \\ 
& =& - \frac{1}{\sqrt{6}} 
\left( 2 \downarrow \downarrow \uparrow - 
\downarrow \uparrow   \downarrow -
\uparrow   \downarrow \downarrow \right).
\ea
Later we will write the spin states in a covariant form.
In the following we  suppress 
the isospin projection index from $\phi_I^{0,1}$
[$+1/2$ as in the proton, 
and  $-1/2$ as in the neutron].

\subsubsection{Nucleon wave function}

With the previous 
notation we write the 
nucleon wave function for
spin projection $s=\pm \sfrac{1}{2}$ as
\be
\Psi_N= \frac{1}{\sqrt{2}} 
\left\{
\phi_I^0 \ket{\sfrac{1}{2}, s}_\rho  +
\phi_I^1 \ket{\sfrac{1}{2}, s}_\lambda 
\right\} \psi_N,
\label{eqPsiN1}
\ee
where $\psi_N$ is a scalar wave function for the quark momentum distribution. 
See Ref.~\cite{Nucleon} for details about the 
nucleon wave function.

\subsubsection{$N(1535)$ non relativistic wave function}

The $N(1535)$ state has the same 
isospin structure of the nucleon.
For  the orbital angular momentum excitation of that state, we consider $L=1$.
We have then the form
\be
\Psi_{S11}= 
\frac{ {\cal N} }{\sqrt{2}}
\left\{
\phi_I^0 X_\rho  -
\phi_I^1 X_\lambda 
\right\} \psi_{S11},
\label{eqPsiS11b}
\ee
with the states $X_\rho$ and $X_\lambda$, 
functions of   $s=\pm \sfrac{1}{2}$,
to be defined next.
The minus sign in the $\lambda$-type term  
is included to ensure the orthogonality 
with the nucleon wave function (\ref{eqPsiN1}).
By construction, $\Psi_{S11}$ is anti-symmetric 
\cite{Aznauryan08,Isgur77,Chiang03}.
The normalization constant ${\cal N}$  
will be determined later.

Here we take the $N(1535)$ state to be composed by states 
with core spin 1/2 only.
The same approximation is 
used in Ref.~\cite{Aznauryan08}.
Alternative models, like the classical Karl-Isgur model
\cite{Capstick95,Isgur77}, where 
the baryons are confined quarks 
with color hyperfine interaction, 
describe $N(1535)$ as a  mixture 
of states with core spin 1/2 and 3/2 
\cite{Capstick95,Isgur77,Chiang03}.

The states $X_\rho$ and $X_\lambda$ 
are combinations of the three quark system mixed-symmetric states,
with total spin 1/2 ($\chi^\rho$ or $\chi^\lambda$) 
and orbital angular momentum $L=1$.
Those states are the direct product 
of orbital angular momentum $L=1$ with 
a spin $1/2$ state.
Considering the product for the projection $s$,
one has, for the mixed-symmetric states $X_\rho$ :
\ba
X_\rho\left( s \right)=
\sqrt{4\pi}
\sum_{m}
\braket{1 m; \sfrac{1}{2}, +\sfrac{1}{2}}{\sfrac{1}{2}, s}
Y_{1,m}(\hat k_\lambda)  
\ket{\sfrac{1}{2}, s-m}_\rho. \nonumber \\ 
\ea
The factor $\sqrt{4\pi}$ was introduced 
by convenience.
Possible
terms in $Y_{1m}(\hat k_\rho)$, associated 
with P states in the diquark, are not considered here.
This corresponds to a pointlike approximation 
for the diquark ($k_\rho \equiv 0$).
Note that the inclusion of structure in the 
diquark, which demands that a dependence 
of the scalar wave function in  $k_\rho$ 
is included
in general \cite{Aznauryan08,Capstick95,Isgur77,Chiang03}.
Here, the  pointlike diquark 
is a first approximation.

As for the $X_\lambda(s)$ states, one has
\ba
X_\lambda\left( s \right)= 
\sqrt{4\pi}
\sum_{m}
\braket{1 m; \sfrac{1}{2}, +\sfrac{1}{2}}{\sfrac{1}{2}, s}
Y_{1,m}(\hat k_\lambda)  
\ket{\sfrac{1}{2}, s-m}_\lambda. \nonumber \\ 
\ea
Once again, we took a pointlike diquark 
[no terms in $Y_{1m}(\hat k_\rho)$].

The spherical harmonics allows us to write the angular momentum states as
\ba
& &
|k_\lambda| Y_{1,+1}(\hat k_\lambda)= 
\sqrt{\frac{3}{4 \pi}} k_{\lambda +} \\
& &
|k_\lambda| Y_{1,\;0}(\hat k_\lambda)= 
\sqrt{\frac{3}{4 \pi}} k_{\lambda 0} \\
& &
|k_\lambda| Y_{1,-1}(\hat k_\lambda)= 
\sqrt{\frac{3}{4 \pi}} k_{\lambda -} 
\ea
where $k_{\lambda 0}= k_{\lambda z}$, and
\be
k_{\lambda \pm}= \mp
\frac{1}{\sqrt{2}} (k_{\lambda x} \pm i
k_{\lambda y}).
\ee 

Replacing the Clebsch-Gordan  coefficients, and
using the compact notation $\pm$ to represent $\pm 1/2$,
one obtains:
\ba
& &
X_\rho\left( \pm \right) = 
\mp N
\left\{
k_{\lambda 0} 
\ket{\sfrac{1}{2}, \pm }_\rho   -
\sqrt{2}
k_{\lambda \pm} 
\ket{\sfrac{1}{2}, \mp}_\rho
\right\}. \nonumber \\
& &
X_\lambda\left( \pm \right) = 
\mp N
\left\{  k_{\lambda 0} 
\ket{\sfrac{1}{2}, \pm }_\lambda   -
\sqrt{2}
 k_{\lambda \pm } 
\ket{\sfrac{1}{2}, \mp}_\lambda
\right\}, \nonumber \\
& & \label{eqX12}
\ea
where $N=1/{|k_\lambda|}$. 
These expressions reproduce the results from Refs.~\cite{Aznauryan08,Chiang03},
in the pointlike diquark limit. 
In that case only the normalization factor differs.

\subsubsection{Normalization}

The normalization of $\Psi_{S11}$
is given by Eq.~(\ref{eqPsiS11b}) [non relativistic form].
Details associated with parity 
will be discussed later in the relativistic generalization.

The wave function (\ref{eqPsiS11b}) must
be normalized in order to reproduce the $N(1535)$ charge:
\ba
Q_{S11} &=& \sum_{\Lambda} \int_k 
\Psi_{S11}^\dagger (\bar P,k) (3 j_1) \Psi_{S11}(\bar P,k)
\nonumber \\
&=& \sfrac{1}{2}(1 + \tau_3).
\label{eqQS11}
\ea
where $\Lambda$ represents the scalar component ($s$) 
and the vectorial component (polarizations $\lambda_D=0,\pm 1$)
of the intermediate diquark, and 
$\bar P=(M_S,0,0,0)$ 
[the momentum configuration correspondent to the rest frame].

The operator $3 j_1= \sfrac{1}{2} + \sfrac{3}{2} \tau_3$ is 
the quark charge operator, where $\tau_3$ acts on the $N(1535)$
isospin states.
In the following we use the notation introduced in the
paper with calculations for the nucleon \cite{Nucleon}.
We project the states into isospin components,
for the case $Q^2=0$, according to 
\ba
& &
j_1 \to \left(\phi_I^0 \right)^\dagger j_1 \phi_I^0 = 
\frac{1}{6} + \frac{1}{2} \tau_3 \\
& &
j_3 \equiv \left(\phi_I^1 \right)^\dagger j_1 \phi_I^1 =
\frac{1}{6} - \frac{1}{6} \tau_3.
\ea
Then considering (\ref{eqX12}),  
one can write
\be
Q_{S11}=
\frac{1}{2}{\cal N}^2
\int_k |\psi_{S11}(\bar P,k)|^2 
\left[ 3 j_1 X_\rho^\dagger X_\rho + 
3 j_3 X_\lambda^\dagger X_\lambda
\right].
\ee
From Eqs.~(\ref{eqX12}), and   
working the spin algebra, for \mbox{$s= \pm 1/2$}, 
one concludes that
\ba
X_\rho^\dagger(s) X_\rho(s) &=& 1 \\
X_\lambda^\dagger(s) X_\lambda(s) &=& 1.
\ea
Then
\ba
Q_{S11}&=& 
{\cal N}^2
\frac{3}{2} 
\left( j_1 + j_3 \right)
\int_k |\psi_{S11}(\bar P,k)|^2 \nonumber \\
 &=& 
\frac{1}{2}(1+ \tau_3)
{\cal N}^2 \int_k |\psi_{S11}(\bar P,k)|^2,
\ea
because $3 (j_1 + j_3)= (1+ \tau_3)$.
Choosing 
\be
\int_k |\psi_{S11}(\bar P,k)|^2=1,
\ee
and one reproduces the $N(1535)$ charge 
(\ref{eqQS11}), if we set ${\cal N}=1$.

\subsection{Relativistic generalization}

The relativistic generalization of 
$k_\lambda$ is the diquark three momentum 
in the rest frame $\tilde k$:
\be
k_\lambda \to 
\tilde k= k - \frac{P \cdot k}{M_S^2} P,
\ee
where $P$ is the $N(1535)$ momentum.
The factor between $k_\lambda$ and $k$ 
from Eq.~(\ref{eqKlambda}) was dropped. 
That factor is included into the normalization of the states.
As $\tilde k^2= -{\bf k}^2$, 
where ${\bf k}$ is the quark three momentum 
in the rest frame, one has
\be
|k_\lambda| \to \sqrt{-\tilde k^2}.
\ee

The diquark  momentum components 
can also be defined in terms of the 
diquark polarization vectors:
\ba
& &
k_{\lambda 0} \to - \tilde k \cdot \varepsilon_P (0) \nonumber \\
& &
k_{\lambda +} \to - \tilde k \cdot \varepsilon_P (+) \nonumber \\
& &
k_{\lambda -} \to - \tilde k \cdot \varepsilon_P (-). 
\label{eqKE}
\ea
In the following we will use $\varepsilon_0$ 
and $\varepsilon_\pm$ for, respectively, 
$\varepsilon_P(0)$ and  $\varepsilon_P(\pm)$.

To obtain the relativistic generalization of 
Eq.~(\ref{eqPsiS11b}), one has to write 
the relativistic generalization of the spin 
states states $\ket{\sfrac{1}{2}, s}_{\rho,\lambda}$.
We use the the covariant 
generalizations,
as 
in the applications to the nucleon system~\cite{Nucleon,NDelta}:
\ba
& &
\ket{ \frac{1}{2}, s}_{\rho} \to \varepsilon^s u(P,s) 
\label{eqPhiR} \\
& &
\ket{ \frac{1}{2}, s}_{\lambda} \to 
-\left(\varepsilon_P^\ast \right)_\alpha U^\alpha(P,s),
\label{eqPhiL}
\ea
where
\ba
U^\alpha(P,s)=
\frac{1}{\sqrt{3}}
\gamma_5 \left(\gamma^\alpha -
\frac{P^\alpha}{M} \right) u(P,s).
\label{eqUa}
\ea
In the previous equations $\varepsilon^s$ is the 
scalar diquark polarization 
$\varepsilon^s = \frac{1}{\sqrt{2}}
( \uparrow \downarrow -\downarrow \uparrow)$
and $\varepsilon_P$ the spin 1 polarization vector 
in the fixed-axis polarization base \cite{Nucleon,FixedAxis,NDelta}.
As $\varepsilon^s$ is a scalar it can replaced by 1
in the wave functions of the nucleon and $N(1535)$.

The expressions for $X_\rho$ and $X_\lambda$ from 
Eqs.~(\ref{eqX12}) can now 
be written in a relativistic form 
using Eqs.~(\ref{eqKE}), (\ref{eqPhiR}) and (\ref{eqPhiL}).
The states $X_\rho$ and $X_\lambda$ are then 
functions of $P$, $k$ (or $P$ and $\tilde k$)
and $s$, but the momentum dependence 
will be suppressed in our the 
notation.
To avoid the dependence of the spin polarization in 
Eqs.~(\ref{eqX12}) on the normalization 
factor,
in the relativistic generalization we replace 
the factor $\mp$ by $-1$, obtaining
a unique expression for both polarizations. 
The final expression is then
\ba
\hspace{-1cm}
X_\rho (\pm) &=&
N \left[
(\tilde k \cdot \varepsilon_0) u_S(\pm) 
 - \sqrt{2} (\tilde k \cdot \varepsilon_\pm) u_S(\mp) 
\right] 
\label{eqXR}
\\
\hspace{-1cm}
 X_\lambda (\pm) &=&
N
\left[ 
- (\tilde k \cdot \varepsilon_0) 
(\varepsilon_P^\ast)_\alpha U_S^\alpha(\pm) 
 \right. \nonumber \\
&& \left. +
\sqrt{2} (\tilde k \cdot \varepsilon_\pm) 
(\varepsilon_P^\ast)_\alpha U_S^\alpha(\mp) \right],
\label{eqXL}
\ea
where we include the sub-index $S$ to label the $N(1535)$ states.
In the previous equations we have replaced 
the non relativistic constant $N= 1/|{\bf k}|$ 
by a new constant such that $|N|=1/\sqrt{- \tilde k^2}$.
The absolute value of $N$ will be fixed 
by the comparison with the experimental data and is
discussed later.

\subsection{$N(1535)$ relativistic wave function}

The final expression for the 
covariant $N(1535)$ wave function, with respect to spin 
flavor, orbital angular momentum and parity,
is then
\be
\Psi_{S11}(P,k)
=\frac{1}{\sqrt{2}}
\gamma_5
\left[ 
\phi_I^0 X_\rho  -
\phi_I^1 X_\lambda 
\right] \psi_{S11}(P,k).
\label{eqPsiRel1}
\ee
The operator $\gamma_5$ was introduced to 
represent the parity of the state.
%
The scalar wave functions were discussed in
the main text (see Sect.~\ref{secWF}).
Equation~(\ref{eqPsiRel1}) reproduces also the $N(1535)$ charge.
With the form (\ref{eqPsiRel1}), one has 
\be
\not \! P \Psi_{S11}= -M_S \Psi_{S11}.
\label{eqDirac}
\ee
This relation (with the minus sign) 
is a consequence of the introduction 
of the operator $\gamma_5$ required by parity.
Note that the $N(1535)$ Dirac equation (\ref{eqDirac})
differs from the equations corresponding 
to the previous applications  
of the spectator quark model 
\cite{Nucleon,NDelta,NDeltaD,LatticeD,Roper}
(nucleon, $\Delta$, and Roper).

In the following we will use
\be
\Psi_{S11}(P,k)
=\frac{1}{\sqrt{2}}
\left[ 
\phi_I^0 \Phi_\rho  -
\phi_I^1 \Phi_\lambda
\right] \psi_{S11}(P,k),
\label{eqPsiRel2}
\ee
where $\Phi_\rho= \gamma_5 X_\rho$ and 
$\Phi_\lambda= \gamma_5 X_\lambda$.

\section{Transition current}
\label{apCurrent}

In this appendix we calculate the
electromagnetic transition current defined by 
Eq.~(\ref{eqJI1}),
using the nucleon and $N(1535)$ 
wave functions given by Eqs.~(\ref{eqPsiN})
and (\ref{eqPsiS11}).

\subsection{$N(1535)$ states}

The $N(1535)$ wave function is given by 
Eq.~(\ref{eqPsiRel2}) with the spin states 
defined by (\ref{eqXR})-(\ref{eqXL}).
From the relations
$ \overline \Phi_\rho \equiv \Phi_\rho^\dagger \gamma^0= 
- \overline X_\rho \gamma_5$ and
$ \overline \Phi_\lambda \equiv \Phi_\lambda^\dagger \gamma^0= 
- \overline X_\lambda \gamma_5$, 
one can write
\ba
& &
\overline \Phi_\rho(\pm) =
- N \left[
(\varepsilon_0 \cdot \tilde k) \bar u_S(\pm) -
\sqrt{2} (\varepsilon_\pm^\ast \cdot \tilde k) \bar u_S(\mp) 
\right] \gamma_5 \nonumber \\
& &
\overline \Phi_\lambda(\pm) =
N \left[
(\varepsilon_0 \cdot \tilde k) \varepsilon_\alpha 
\overline U_S^\alpha (\pm) -
\sqrt{2} (\varepsilon_\pm^\ast \cdot \tilde k)  
 \varepsilon_\alpha
\overline U_S^\alpha (\mp)
\right] \gamma_5. \nonumber \\
& & 
\ea
In the previous equations 
\be
\overline U_S^\alpha = -\frac{1}{\sqrt{3}} 
\bar u_S \left(\gamma^\alpha - \frac{P^\alpha}{M_S}\right) \gamma_5. 
\ee

\subsection{Properties of the states}

To reduce the transition current to the standard form
one uses the properties of the nucleon and 
$N(1535)$ spin states $U_S^\alpha$, $u_S$, $U^\alpha $ and $u$:
\ba
& &\not \! P_- u(P_-)=M u(P_-) \nonumber \\
& &\not \! P_-  U^\alpha(P_-)= M U^\alpha(P_-)  
\nonumber \\
& &
\not \! P_+ u_S(P_+)=M_S u_S(P_+) \nonumber \\
& &
\not \! P_+  U_S^\alpha(P_+)= M_S U_S^\alpha(P_+).  
\ea
Also
\ba
& &
(P_+)_\alpha U_S^\alpha=0 \\
& &
(P_-)_\alpha  U^\alpha=0.
\ea

\subsection{Integration in $k$}

In the following we consider the symmetries in 
the $k$ integration.
The evaluation of the transition current 
requires the determination of the
integrals
\be
{\cal I}_{\lambda^\prime}=
\int_k N (\varepsilon_{\lambda^\prime} \cdot \tilde k) \psi_{S11} \psi_N,
\ee
where $\lambda^\prime= 0,\pm$.
It is easy to prove that
\be
{\cal I}_\pm =0,
\ee
for any value of $Q^2$. 
The demonstration is 
trivial in the $N(1535)$ rest frame, since 
the product of wave functions can be written 
as a function of ${\bf k}^2$ and $k_z$.
Then $\int_k N k_x \psi_{S11} \psi_N= \int_k N k_y \psi_{S11} \psi_N=0$,
because the integrand function is odd in the 
integration variables $k_{x,y}$.
Then only  ${\cal I}_0$ survives the 
$k$ integration for a given $Q^2$.
The case $Q^2=0$ will be discussed in 
Appendix \ref{apINT}.
The important point here, is that in the final state rest frame we have to keep 
in the wave function
only the terms in $\tilde k \cdot \varepsilon_0= - k_z$.

\subsection{Current matrix elements}

Considering the expression for 
the spin states and by performing the integral for the current, one obtains, 
for arbitrary  (initial and final) 
spin projections:
\ba
& &\overline \Phi_\rho \hat \gamma^\mu \phi_S^0=
N ( \varepsilon_0 \cdot \tilde k) \bar u_S  \hat \gamma^\mu 
\gamma_5 u \label{eqPhi1A}\\
& &\overline \Phi_\rho 
  \frac{i \sigma^{\mu \nu} q_\nu}{2M}
 \phi_S^0=
- N ( \varepsilon_0 \cdot \tilde k) \bar u_S  
 \frac{i \sigma^{\mu \nu} q_\nu}{2M}
\gamma_5 u \label{eqPhi1B}\\
& &
\overline \Phi_\lambda  \hat \gamma^\mu \phi_S^1=
N ( \varepsilon_0 \cdot \tilde k) \left[
\overline U_S^\alpha   \hat \gamma^\mu 
\gamma_5 U^\beta \right] \Delta_{\alpha \beta}
\label{eqPhi2A} \\
& &
\overline \Phi_\lambda  
  \frac{i \sigma^{\mu \nu} q_\nu}{2M} 
 \phi_S^1=
- N ( \varepsilon_0 \cdot \tilde k) \left[
\overline U_S^\alpha     \frac{i \sigma^{\mu \nu} q_\nu}{2M}
\gamma_5 U^\beta \right] \Delta_{\alpha \beta}. \nonumber \\
& & \label{eqPhi2B}
\ea
In the previous equations 
$\Delta_{\alpha \beta}$ is given by Eq.~(\ref{eqDelta}).
For the terms $i \sigma^{\mu \nu} q_\nu$ 
one can use the generalized Gordon identity:
\be
i \sigma^{\mu \nu} q_\nu=
\not \!P_+ \gamma^\mu + \gamma^\mu \not \! P_- -
(P_+ + P_-)^\mu.
\ee

\subsection{Spin algebra}

The following relations holds when multiplied by $\Delta_{\alpha \beta}$:
\ba
& &
\overline U_S^\alpha \hat \gamma^\mu \gamma_5 U^\beta=
-\frac{1}{3} 
\bar u_S \gamma^\alpha \hat \gamma^\mu 
\gamma^\beta \gamma_5 u\\
& & \overline U_S^\alpha \frac{i \sigma^{\mu \nu} q_\nu}{2M} 
\gamma_5 U^\beta=
\frac{1}{3} 
\bar u_S \gamma^\alpha \frac{i \sigma^{\mu \nu} q_\nu}{2M}
\gamma^\beta \gamma_5 u.
\ea
Considering the results:
\ba
& &
\left[
\gamma^\alpha  \gamma^\mu \gamma^\beta \gamma_5\right] 
\Delta_{\alpha \beta}= \gamma^\mu \gamma_5 \\
& &
\left[
\gamma^\alpha \gamma^\beta \gamma_5\right] 
\Delta_{\alpha \beta}= -\gamma_5, 
\ea
one obtains
\ba
& &
\overline U_S^\alpha \hat \gamma^\mu \gamma_5 U^\beta=
\frac{1}{3} 
\bar u_S  \hat \gamma^\mu \gamma_5 u\\
& &
\overline U_S^\alpha \frac{i \sigma^{\mu \nu} q_\nu}{2M} 
\gamma_5 U^\beta=
-\frac{1}{3} 
\bar u_S  \frac{i \sigma^{\mu \nu} q_\nu}{2M} \gamma_5 u.
\ea

\subsection{Final expressions}

Using the formulas of the
previous section one can write 
the result of the integration in $k$ 
for (\ref{eqPhi1A})-(\ref{eqPhi2B})
including also
$\psi_{S11}(P_+,k) \psi_N(P_-,k)$.
In the integration the relations with 
$(\varepsilon_0 \cdot \tilde k)$ 
are replaced  by ${\cal I}_0$,
defined by Eq.~(\ref{eqInt0}).
Then 
\ba
& &
\hspace{-1.3cm}
\int_k \left[
\overline \Phi_\rho \hat \gamma^\mu \phi_S^0 \right] \!
\psi_{S11} \psi_N
=
{\cal I}_0 \bar u_S  \hat \gamma^\mu 
\gamma_5 u \\
& &
\hspace{-1.3cm}
\int_k 
\left[
\overline \Phi_\rho 
  \frac{i \sigma^{\mu \nu} q_\nu}{2M}
 \phi_S^0 \right] 
\!\psi_{S11} \psi_N =
- {\cal I}_0 \bar u_S  
 \frac{i \sigma^{\mu \nu} q_\nu}{2M}
\gamma_5 u \\
& &
\hspace{-1.3cm}
\int_k
\left[ 
\overline \Phi_\lambda  \hat \gamma^\mu \phi_S^1\right]
\!\psi_{S11}\psi_N=
-\frac{1}{3}  {\cal I}_0
\bar u_S  \hat \gamma^\mu \gamma_5 u\\
& &
\hspace{-1.3cm}
\int_k
\left[
\overline \Phi_\lambda  
  \frac{i \sigma^{\mu \nu} q_\nu}{2M}  \phi_S^1\right] \! \psi_{S11}\psi_N=
\frac{1}{3} {\cal I}_0
\bar u_S  \frac{i \sigma^{\mu \nu} q_\nu}{2M} \gamma_5 u.
\ea
Replacing the previous results in the 
expression for the current, we obtain
\ba
J^\mu &=& + \frac{1}{2}(3 j_1+ j_3) 
{\cal I}_0 \bar u_S \hat \gamma^\mu  \gamma_5 u \nonumber \\
& &- \frac{1}{2}(3 j_2- j_4) 
{\cal I}_0 \bar u_S 
\frac{i \sigma^{\mu \nu} q_\nu}{2M} 
  \gamma_5 u. 
\ea

The previous current defines the electromagnetic 
transition form factors given by Eqs.~(\ref{eqF1})-(\ref{eqF2}).
The sign of the normalization constant $N$ 
with magnitude $|N|=\sfrac{1}{\sqrt{-\tilde k^2}}$ 
has to be fixed by the experimental data.
As the data for $F_1^\ast$ is negative near $Q^2=0$, 
we choose
\ba
N= - \frac{1}{\sqrt{-\tilde k^2}}.
\label{eqN}
\ea

\section{Overlap integral ${\cal I}_0$}
\label{apINT}

In this appendix we consider the integral 
of Eq.~(\ref{eqInt0}):
\be
{\cal I}_0 = 
\int_k N(\epsilon_0 \cdot \tilde k) \psi_{S11}(P_+,k) \psi_N(P_-,k).
\label{eqInt0a}
\ee
First, we derive an analytical expression for ${\cal I}_0$,
next we explore the limit cases.

\subsection{Analytical expression}

Consider the expression for the overlap integral (\ref{eqInt0a}),
in the $N(1535)$ rest frame
 \be
{\cal I}_0 = 
\int_k \frac{k_z}{|{\bf k}|} \psi_{S11}(P_+,k) \psi_N(P_-,k),
\label{eqInt0b}
\ee
where we used $\tilde \epsilon_0 \cdot \tilde k= - k_z$ 
and Eq.~(\ref{eqN}).
In the same frame one has 
$P_+=(M_S,0,0,0)$, $P_-=(E,0,0,-|{\bf q}|)$ 
and \mbox{$q=(\omega,0,0,|{\bf q}|)$}, 
with $\omega=M_S-E$ and
\ba
& &
\hspace{-1cm}
E=\frac{M_S^2+M^2+Q^2}{2M_S}.
\ea
In this case $\psi_{S11}$ is 
independent of the 
azimuthal angle and we can write, using $k_z = kz$:
\ba
{\cal I}_0= \int_0^{+\infty}
\frac{k^2 dk}{(2\pi)^2 2E_D} 
\psi_{S11}(P_+ \cdot k) I_z,
\label{eqInt02}
\ea
where
\be
I_z = \int_{-1}^1 \left[z \psi_N(P_- \cdot k) \right]dz.  
\ee
In the previous equation we use the simplified 
notation for the arguments of the wave functions, 
since they can be represented as a scalar function 
of $P_\pm \cdot k$:
\ba
P_+ \cdot k= M_S E_D, \hspace{.5cm}
P_- \cdot k= E E_D + k_z |{\bf q}|.
\ea

The separation of the function that depends on $z$ 
as $\psi_{N}$ from the ones depending 
only of $k$, as $\psi_{S11}$
is possible because in the $N(1535)$ rest frame 
$P_+ \cdot k$ is angle independent.
As for $\psi_{N}$, it is represented by the simple 
analytical form (\ref{eqPsiNS}). In these 
conditions one can evaluate $I_z$ 
analytically using simple integration techniques.
The result is 
\ba
I_z &=& 
\frac{N_0}{m_D}
\left(\frac{M m_D}{2 k |{\bf q}|} \right)^2
\frac{1}{\beta_2 -\beta_1}
\times \nonumber \\
& &
\left[
\bar \beta_2 G_2 (k,|{\bf q}|) - \bar \beta_1 G_1 (k,|{\bf q}|)
\right]
\label{eqIz}
\ea
where
\ba
& &
\bar \beta_i= (\beta_i -2) + 2 \frac{E E_D}{M m_D} \\
& &
G_i(k, |{\bf q}|)=
\log \left| \frac{ \bar \beta_i +   2 \frac{k |{\bf q}|}{M m_D}  }{ 
\bar \beta_i -   2 \frac{k |{\bf q}|}{M m_D}}
\right|.
\ea

To obtain the final expression one has to perform 
the integration in $k$.

\subsection{${\cal I}_0$ in the limit $|{\bf q}| \to 0$}

The expression obtained for $I_z$ from Eq.~(\ref{eqIz}) 
does not help us to explore the limit $Q^2 \to 0$.
To have a clearer idea of the $Q^2$ or $|{\bf q}|$ 
dependence one considers 
the case $|{\bf q}| \to 0$.
In that limit one case use 
\ba
\log \left| \frac{1+x}{1-x} \right|=
2 x + \frac{2}{3} x^3 + {\cal O}(x^5),
\ea
to simplify $I_z$.
Using the previous equation with 
\ba
x= \frac{ 2k }{M m_D} \frac{|{\bf q}|}{\bar \beta_i},
\ea
one can conclude that
\ba
I_z= - \frac{4}{3} \frac{N_0}{m_D} 
\left( \frac{k}{M m_D} \right) 
\frac{ \bar \beta_1 + \bar \beta_2}{
\bar \beta_1^2 \bar \beta_2^2} |{\bf q}|.
\ea

With this relation we prove that
\be
{\cal I}_0 (Q^2) \propto |{\bf q}|,
\ee
for small $|{\bf q}|$.

\subsection{Two different limits}

For the equal mass case ($M_S=M$), where 
\be
|{\bf q}|= \sqrt{1+ \tau} Q,
\ee 
using $\tau= \sfrac{Q^2}{(M_S+M)^2}\equiv \sfrac{Q^2}{4M^2}$,
one can conclude that 
\be
{\cal I}_0 (Q^2) \propto Q, 
\ee
implying that $F_1^\ast(Q^2),F_2^\ast(Q^2) \propto Q$ as $Q^2 \to 0$.
This dependence is atypical and unexpected.
Recall that the nucleon to Roper 
form factors vanish for $Q^2 \to 0$ with the power $Q^2$
(in that specific case, independently of the mass difference).

In $Q^2=0$ limit,
and in the unequal mass case, one has
\be
|{\bf q}|=|{\bf q}|_0
\equiv \frac{M_S^2-M^2}{2M_S}.
\ee
In this situation one concludes that
\ba
{\cal I}_0 (0) \propto \frac{M_S^2-M^2}{2M_S}.
\ea 
This last result implies that the $S_{1/2}(Q^2)$ amplitude diverges 
for $M_S \ne M$.
As that amplitude scales with $1/Q^2$,  for $Q^2 \to 0$,
if ${\cal I}_0 (0) \ne0$,  
the amplitude diverges for $Q^2 \to 0$. 
For the form factors $F_1^\ast$ 
and $F_2^\ast$ there is no divergence for $Q^2 \to 0$.

\end{document}